\documentclass[english,twocolumn]{aastex61}
\usepackage{lmodern}
\usepackage[T1]{fontenc}
\usepackage[latin9]{inputenc}
\usepackage{color}
\usepackage{booktabs}
\usepackage{amsbsy}
\usepackage{amstext}
\usepackage{graphicx}

\makeatletter

\providecommand{\tabularnewline}{\\}

\usepackage{amsmath}

\shorttitle{Plasmoid Instability and Fast Reconnection}
\shortauthors{Huang et al.}

\makeatother

\usepackage{babel}
\begin{document}

\title{Plasmoid Instability in Evolving Current Sheets and Onset of Fast
Reconnection}

\correspondingauthor{Yi-Min Huang} \email{yiminh@princeton.edu} 

\author{Yi-Min Huang}

\affil{Department of Astrophysical Sciences, Princeton Plasma Physics Laboratory,
Princeton Center for Heliophysics, and Max Planck/Princeton Center
for Plasma Physics, Princeton University, Princeton, New Jersey 08543,
USA}

\author{Luca Comisso}

\affil{Department of Astrophysical Sciences, Princeton Plasma Physics Laboratory,
Princeton Center for Heliophysics, and Max Planck/Princeton Center
for Plasma Physics, Princeton University, Princeton, New Jersey 08543,
USA}

\author{A. Bhattacharjee}

\affil{Department of Astrophysical Sciences, Princeton Plasma Physics Laboratory,
Princeton Center for Heliophysics, and Max Planck/Princeton Center
for Plasma Physics, Princeton University, Princeton, New Jersey 08543,
USA}
\begin{abstract}
The scaling of plasmoid instability maximum linear growth rate with
respect to Lundquist number $S$ in a Sweet-Parker current sheet,
$\gamma_{max}\sim S^{1/4}$, indicates that at high $S$, the current
sheet will break apart before it approaches the Sweet-Parker width.
Therefore, a proper description for the onset of the plasmoid instability
must incorporate the evolving process of the current sheet. We carry
out a series of two-dimensional simulations and develop diagnostics
to separate fluctuations from an evolving background. It is found
that the fluctuation amplitude starts to grow only when the linear
growth rate is sufficiently large $(\gamma_{max}\tau_{A}>O(1))$ to
overcome advection loss and the stretching effect due to the outflow.
The linear growth rate continues to rise until the sizes of plasmoids
become comparable to the inner layer width of the tearing mode. At
this point the current sheet is disrupted and the instability enters
the early nonlinear regime. The growth rate suddenly decreases, but
the reconnection rate starts to rise rapidly, indicating that current
sheet disruption triggers the onset of fast reconnection.  We identify
important time scales of the instability development, as well as scalings
for the linear growth rate, current sheet width, and dominant wavenumber
at disruption. These scalings depend on not only the Lundquist number,
but also the noise amplitude. A phenomenological model that reproduces
scalings from simulation results is proposed. The model incorporates
the effect of reconnection outflow, which is crucial for yielding
a critical Lundquist number $S_{c}$ below which disruption does not
occur. The critical Lundquist number $S_{c}$ is not a constant value
but has a weak dependence on the noise amplitude. 
\end{abstract}

\section{Introduction}

Most of the visible universe is in plasma state, where the presence
of magnetic field is ubiquitous. At large spatial and slow temporal
scales relative to kinetic scales, the dynamics of highly conducting
magnetized plasmas is governed by ideal magnetohydrodynamics (MHD)
equations. A topological constraint imposed by ideal MHD is the frozen-in
condition, which implies the preservation of field line connectivity
among fluid elements \citep{Alfven1943}. Magnetic reconnection is
a fundamental process that allows plasmas to break free from the frozen-in
constraint, thereby enabling the conversion of magnetic energy to
plasma energy. Magnetic reconnection is generally believed to be the
underlying mechanism that powers explosive events in nature, such
as magnetospheric substorms, solar flares, coronal mass ejections
(CME), gamma-ray bursts, and sawtooth crashes in fusion plasmas \citep[see][for recent reviews]{Biskamp2000,PriestF2000,ZweibelY2009,YamadaKJ2010,ZweibelY2016}.

Magnetic reconnection events in nature are often preceded by an extended
quiescent period and then an impulsive ``onset'' phase. During the
onset phase the reconnection rate increases abruptly, and subsequently
reconnection proceeds rapidly after the onset \citep{Bhattacharjee2004,CassakSD2005}.
Accordingly, a successful theory of reconnection must account for
not only the observed fast reconnection rate, but also the impulsive
onset. The classical Sweet-Parker theory \citep{Parker1957,Sweet1958a}
based on resistive MHD fails to accomplish both tasks. The Sweet-Parker
theory assumes that reconnection takes place in an extended current
sheet of length $L$, which is determined by the global configuration.
The governing dimensionless parameter of the system is the Lundquist
number $S\equiv V_{A}L/\eta$, where $V_{A}$ is the Alfv\'en speed
of the reconnecting magnetic field and $\eta$ is the magnetic diffusivity.
The theory predicts that the width of the current sheet scales as
$a_{SP}\simeq LS^{-1/2}$, and from conservation of mass the reconnection
rate also scales as $S^{-1/2}$. Because the Lundquist number is typically
very large in astrophysical systems (\emph{e.g.} $S\sim10^{12}\text{--}10^{14}$
in solar corona), the Sweet-Parker reconnection rate is often too
slow to account for observations. Furthermore, because the Sweet-Parker
theory is inherently a steady-state theory, it also cannot explain
the impulsive onset. 

In recent years, significant progress has been made in understanding
the fast reconnection rate. In particular, it is now well-established
that the Sweet-Parker theory is not the correct description for resistive
MHD reconnection in the high-Lundquist-number regime, because the
reconnection layer is unstable to the plasmoid instability \citep{Biskamp1993},
which is a tearing-type instability \citep{FurthKR1963}. For two-dimensional
(2D) reconnection, the elongated current sheet in the Sweet-Parker
theory is replaced by a chain of plasmoids (or magnetic islands) and
fragmented thin current sheets. In fully developed nonlinear stage,
the reconnection rate becomes nearly independent of the Lundquist
number \citep{Lapenta2008,BhattacharjeeHYR2009,CassakSD2009,HuangB2010,UzdenskyLS2010,LoureiroSSU2012,NiZHLM2012,HuangB2013,Takamoto2013,ComissoGF2015,LoureiroU2016}.
Furthermore, if the fragmented current sheets become thinner than
kinetic scales such as ion inertial length $d_{i}$ or ion gyroradius
$\rho_{i}$, even faster Hall/collisionless reconnection can be triggered
\citep{DaughtonRAKYB2009,ShepherdC2010,HuangBS2011,JiD2011}. In three
dimensions (3D), the plasmoid instability can facilitate self-generated
turbulent reconnection \citep{DaughtonRKYABB2011,OishiLCT2015,HuangB2016}.
Plasmoid-mediated reconnection has been studied in a wide range of
contexts, including post-CME current sheets \citep{KarpenAD2012,GuoBH2013,LynchEKG2016},
weakly ionized plasmas \citep{LeakeLLM2012,LeakeLL2013}, embedded
reconnection in a broad outer current sheet \citep{CassakD2009},
asymmetric reconnection \citep{MurphyYSLN2013}, relativistic reconnection
\citep{Takamoto2013,SironiGP2016}, and statistical distribution of
plasmoids \citep{FermoDS2010,UzdenskyLS2010,LoureiroSSU2012,HuangB2012}.

Thus far, magnetic reconnection at the fully developed nonlinear stage
of the plasmoid instability has been the focus of many studies. On
the other hand, onset of fast reconnection facilitated by the plasmoid
instability from the linear regime to nonlinear saturation remains
not well-understood. The linear theory of plasmoid instability assuming
a Sweet-Parker current sheet width $a_{SP}\simeq LS^{-1/2}$ gives
a maximum growth rate $\gamma_{max}$ satisfying the scaling relation
$\gamma_{max}\tau_{A}\sim S^{1/4}$, where $\tau_{A}\equiv L/V_{A}$
is the Alfv\'enic time scale based on the current sheet length, and
the corresponding wavenumber $k_{max}$ of the fastest growing mode
satisfies the scaling relation $k_{max}L\sim S^{3/8}$ \citep{TajimaS1997,LoureiroSC2007}.
These scaling relations indicate that a Sweet-Parker current sheet
in high-$S$ regime is drastically unstable and will generate an enormous
number of plasmoids. However, because a Sweet-Parker sheet must be
realized dynamically over time, the fact that $\gamma_{max}$ diverges
in the asymptotic limit $S\to\infty$ suggests that the current sheet
will break apart before it reaches the Sweet-Parker width \citep{PucciV2014}.
This raises important questions about the condition for current sheet
disruption: when does the current sheet break apart? what is the current
sheet width when it does so? what is the dominant mode that disrupts
the current sheet, and what is the corresponding growth rate? The
answers to these questions have significant implications for the impulsive
onset of fast magnetic reconnection.

The main objective of this paper is to investigate the plasmoid instability
in dynamically evolving current sheets to address these questions,
some of which are currently under debate in recent literature. \citet{PucciV2014}
argued that the maximum linear growth rate at disruption should become
independent of $S$. From the scaling relations $\gamma_{max}\tau_{A}\sim S^{(3\alpha-1)/2}$
and $k_{max}L\sim S^{(5\alpha-1)/4}$ for a general situation where
the current sheet width scales as $a\sim LS^{-\alpha}$ \citep{BhattacharjeeHYR2009,PucciV2014},
Pucci and Velli concluded that $\alpha=1/3$ and proposed the inverse
aspect ratio of the current sheet $a/L=S^{-1/3}$ as the condition
for current sheet disruption; the corresponding maximum linear growth
rate is a fixed value with $\gamma_{max}\tau_{A}=O(1)$ and the wavenumber
satisfies the scaling relation $k_{max}L\sim S^{1/6}$. Two other
recent theoretical studies by \citet{UzdenskyL2016} and \citet{ComissoLHB2016}
came to different conclusions regarding the disruption condition.
\citet{UzdenskyL2016} assumed that the current sheet is essentially
frozen when $\gamma\tau_{dr}=1$, where $\tau_{dr}$ is the evolution
time scale of the current sheet, and concluded that the fastest mode
at that time will be the one that disrupts the current sheet. On the
contrary, \citet{ComissoLHB2016} found that the thinning process
is important until the disruption time. \citet{ComissoLHB2016} employed
a principle of least time to derive analytically the scaling relations
of the current sheet aspect ratio, the linear growth rate, and wavenumber
of the dominant mode at disruption. These scaling relations are not
power laws, with dependences not only on the Lundquist number $S$,
but also on the noise level of the environment. 

This work is distinct from previous studies in that we employ both
direct numerical simulations and theoretical analysis of a phenomenological
model. The corroboration between the two approaches puts our conclusions
in solid footing. Our key findings can be summarized as follows:
\begin{enumerate}
\item The Sweet-Parker current sheet thickness $a_{SP}$ can only be approached
at low to moderately high Lundquist numbers. Therefore, there is a
transition of scaling behavior from low-$S$ to high-$S$ regimes. 
\item The current sheet disruption does not take place when $\gamma_{max}\tau_{A}=O(1)$,
in contrast to the suggestion of \citet{PucciV2014}. In fact, the
overall amplitude of fluctuation only starts to grow when $\gamma_{max}\tau_{A}=O(1)$,
because the linear growth of tearing modes has to overcome the decrease
of fluctuation amplitude due to advection loss and stretching effect
of the outflow. Typically, the maximum linear growth rate satisfies
$\gamma_{max}\tau_{A}\gg1$ at disruption. 
\item Contrary to what is commonly assumed, the dominant mode at disruption
is not the fastest growing mode. Typically the dominant mode has a
wavenumber approximately 3 to 6 times smaller than that of the fastest
growing mode. 
\item The dominant mode and growth rate at disruption cannot be determined
by the Lundquist number $S$ alone, as they also depend on the noise
of the system.\textbf{ }
\item The phenomenological model gives a prediction of the critical Lundquist
number $S_{c}$ for plasmoid instability, below which the instability
cannot grow to a sufficient amplitude to disrupt the current sheet
before it is advected out from the current sheet. Importantly, the
critical Lundquist number $S_{c}$ is not a fixed magnitude but depends
weakly on the noise amplitude. 
\end{enumerate}
This paper is organized as follows. The details of the simulation
setup are given in Section \ref{sec:setup}. Section \ref{sec_Diagnostics}
gives a description of our diagnostics for analyzing fluctuations
in an evolving background and the condition for current sheet disruption.
Section \ref{sec:simulation_results} presents the results from numerical
simulations. In Section \ref{sec:Phenomenological-Model}, we propose
a phenomenological model that reproduces the scalings from simulation
results and provides further insights. In Section \ref{sec:nonlinear},
we discuss the transition from slow to fast reconnection after current
sheet disruption and how the transition may be observed in solar spectroscopy
through line profiles. We conclude and make comparisons with previous
studies in Section \ref{sec:conclusion}.

\section{Simulation Setup}

\label{sec:setup}The governing equations for our simulations are
non-dimensionalized two-dimensional (2D) resistive MHD equations: 

\begin{equation}
\partial_{t}\rho+\nabla\cdot\left(\rho\mathbf{v}\right)=0,\label{eq:1}
\end{equation}
\begin{equation}
\partial_{t}(\rho\mathbf{v})+\nabla\cdot\left(\rho\mathbf{vv}\right)=-\nabla p-\nabla\psi\nabla^{2}\psi+\nu\nabla^{2}(\rho\mathbf{v}),\label{eq:2}
\end{equation}
\begin{equation}
\partial_{t}\psi+\mathbf{v}\cdot\nabla\psi=\eta\nabla^{2}\psi,\label{eq:3}
\end{equation}
where standard notations are used. Here the Cartesian coordinate $y$
is the direction of translational symmetry. The magnetic field $\mathbf{B}$
is related to the flux function $\psi$ via the relation $\mathbf{B}=\nabla\psi\times\hat{\boldsymbol{y}}$
and the electric current density $\mathbf{J}=-\nabla^{2}\psi\hat{\boldsymbol{y}}$.
An isothermal equation of state $p=2\rho T$ with a constant temperature
$T$ is assumed. The numerical algorithm is detailed in \citep{GuzdarDMHL1993},
where derivatives are approximated by a five-point central finite
difference scheme, with a fourth-order numerical dissipation equivalent
to up-wind finite difference added to all equations for numerical
stability. Time stepping is calculated by a trapezoidal leapfrog scheme.
Explicit dissipations are employed through viscosity and resistivity.

We use the same simulation setup of two coalescing magnetic islands
as in a previous study \citep{HuangB2010}. The 2D simulation box
is the domain $(x,z)\in[-1/2,1/2]\times[-1/2,1/2]$. In normalized
units, the initial magnetic field is given by $\mathbf{B}_{0}=\nabla\psi_{0}\times\mathbf{\hat{y}}$,
where $\psi_{0}=\tanh\left(z/h\right)\cos\left(\pi x\right)\sin\left(2\pi z\right)/2\pi$.
The parameter $h$, which is set to $0.01$ for all simulations, determines
the initial current layer width. The initial plasma density $\rho$
is approximately unity, and the plasma temperature $T=3$. The density
profile has a weak nonuniformity such that the initial condition is
approximately force-balanced. The initial peak magnetic field and
Alfv\'en speed are both approximately unity. The plasma beta $\beta\equiv p/B^{2}=2\rho T/B^{2}$
is greater than $6$ everywhere, hence the system is approximately
incompressible. Perfectly conducting and free slipping boundary conditions
are imposed along both $x$ and $z$ directions. Only the upper half
of the domain ($z\ge0$) is simulated, and solutions in the lower
half are inferred by symmetries. We use a uniform grid of 37800 points
along the $x$ direction and a nonuniform grid of 2880 points along
the $z$ direction. The mesh along the $z$ direction is packed to
attain high resolution around $z=0$, where the smallest grid size
is $1.9\times10^{-6}$. The viscosity $\nu$ is set to $10^{-10}$
for all the simulations, and the resistivity $\eta$ is varied from
$2\times10^{-6}$ to $2.5\times10^{-8}$, hence the magnetic Prandtl
number $P_{m}\equiv\nu/\eta\ll1$. The initial velocity is seeded
with a random noise of amplitude $\epsilon$ to trigger the plasmoid
instability. In these simulations, the current sheet half-length is
approximately a constant value $L=0.25$ and the upstream Alfv\'en
speed is approximately $V_{A}=1$. We use these values to define the
Lundquist number $S=LV_{A}/\eta$ of the system. The simulation parameters
and outcomes of diagnostics are summarized in Table \ref{tab:run_table}.

\setlength{\tabcolsep}{4pt} 

\begin{table*}
\begin{centering}
\begin{tabular}{ccccccccccccccc}
\toprule 
Run & $S$ & $\epsilon$ & $t_{g}$ & $\gamma_{max,g}$ & $a_{g}$ & $t_{d}$ & $a_{d}$ & $\delta_{d}$ & $\gamma_{d}$ & $\gamma_{max}$ & $k_{d}$  & $k_{max}$ & $t_{p}$ & $t_{s}$ \tabularnewline
\midrule 
S1 & $\mathrm{1.25e}5$ & $10^{-6}$ & 0.42 & 12.4 & $\mathrm{1.72e-}3$ & 1.24 & $\mathrm{7.47e-}4$ & $\mathrm{2.57e-}4$ & 14.6 & 43.1 & 81.4 & 414 & 1.81 & 2.01\tabularnewline
S2 & $\mathrm{2.5e}5$ & $10^{-6}$ & 0.42 & 9.61 & $\mathrm{1.60e-}3$ & 1.06 & $\mathrm{4.72e-}4$ & $\mathrm{1.69e-}4$ & 29.6 & 60.8 & 92.3 & 617 & 1.28 & 1.42\tabularnewline
S3 & $\mathrm{5.0e}5$ & $10^{-6}$ & $0.45$ & 8.85 & $\mathrm{1.35e-}3$ & $1.00$ & $\mathrm{3.37e-}4$ & $\mathrm{9.98e-}5$ & 44.5 & 71.3 & 157 & 791 & 1.13 & 1.25\tabularnewline
S4 & $\mathrm{1.25e}6$ & $10^{-6}$ & $0.52$ & 9.74 & $\mathrm{9.35e-}4$ & $0.99$ & $\mathrm{2.37e-}4$ & $\mathrm{5.90e-}5$ & 62.1 & 77.7 & 213 & 977 & 1.16 & 1.19\tabularnewline
S5 & $\mathrm{2.5e}6$ & $10^{-6}$ & $0.57$ & 10.3 & $\mathrm{7.16e-}4$ & $1.00$ & $\mathrm{1.83e-}4$ & $\mathrm{3.89e-}5$ & 69.7 & 79.4 & $282$ & 1140 & 1.10 & 1.15\tabularnewline
S6 & $\mathrm{5.0e}6$ & $10^{-6}$ & 0.58 & 8.09 & $\mathrm{6.67e-}4$ & 1.02 & $\mathrm{1.40e-}4$ & $\mathrm{2.82e-}5$ & 74.0 & 83.9 & 287 & 1330 & 1.09 & 1.12\tabularnewline
S7 & $\mathrm{1.0e}7$ & $10^{-6}$ & 0.61 & 7.33 & $\mathrm{5.65e-}4$ & 1.04 & $\mathrm{1.09e-}4$ & $\mathrm{1.88e-}5$ & 77.3 & 86.7 & 374 & 1530 & 1.10 & 1.13\tabularnewline
\midrule
H1 & $\mathrm{1.25e}5$ & $10^{-3}$ & 0.39 & 10.3 & $\mathrm{1.94e-}3$ & 0.87 & $\mathrm{6.77e-}4$ & $\mathrm{2.39e-}4$ & 26.2 & 50.0 & 92.3 & 468 & 1.17 & 1.18\tabularnewline
H2 & $\mathrm{2.5e}5$ & $10^{-3}$ & 0.43 & 10.3 & $\mathrm{1.55e-}3$ & 0.84 & $\mathrm{5.18e-}4$ & $\mathrm{1.51e-}4$ & 36.6 & 52.9 & 136 & 550 & 0.99 & 1.14\tabularnewline
H3 & $\mathrm{5.0e}5$ & $10^{-3}$ & $0.45$ & 8.85 & $\mathrm{1.35e-}3$ & 0.84 & $\mathrm{3.99e-}4$ & $\mathrm{1.03e-}4$ & 43.1 & 53.5 & 169 & 640 & 0.96 & 1.04\tabularnewline
H4 & $\mathrm{1.25e}6$ & $10^{-3}$ & $0.52$ & 9.74 & $\mathrm{9.35e-}4$ & 0.85 & $\mathrm{3.02e-}4$ & $\mathrm{6.66e-}5$ & 45.6 & 53.0 & 185 & 721 & 1.01 & 1.04\tabularnewline
H5 & $\mathrm{2.5e}6$ & $10^{-3}$ & 0.57 & 10.3 & $\mathrm{7.16e-}4$ & 0.88 & $\mathrm{2.33e-}4$ & $\mathrm{4.66e-}5$ & 47.5 & 55.4 & 209 & 839 & 1.01 & 1.05\tabularnewline
H6 & $\mathrm{5.0e}6$ & $10^{-3}$ & 0.58 & 8.09 & $\mathrm{6.67e-}4$ & 0.90 & $\mathrm{1.86e-}4$ & $\mathrm{3.01e-}5$ & 51.0 & 55.1 & 297 & 935 & 1.01 & 1.06\tabularnewline
H7 & $\mathrm{1.0e}7$ & $10^{-3}$ & 0.61 & 7.33 & $\mathrm{5.65e-}4$ & 0.93 & $\mathrm{1.64e-}4$ & $\mathrm{2.09e-}5$ & 52.9 & 56.6 & 346 & 920 & 1.02 & 1.06\tabularnewline
\midrule
N1 & $\mathrm{2.5e}6$ & $10^{-2}$ & 0.57 & 10.3 & $\mathrm{7.16e-}4$ & 0.82 & $\mathrm{2.72e-}4$ & $\mathrm{4.87e-}5$ & 39.5 & 43.9 & 208 & 691 & 0.97 & 1.03\tabularnewline
N2 & $\mathrm{2.5e}6$ & $10^{-5}$ & 0.57 & 10.3 & $\mathrm{7.16e-}4$ & 0.96 & $\mathrm{1.97e-}4$ & $\mathrm{4.11e-}5$ & 64.3 & 71.1 & 256 & 1035 & 1.07 & 1.12\tabularnewline
N3 & $\mathrm{2.5e}6$ & $10^{-9}$ & 0.56 & 9.57 & $\mathrm{7.51e-}4$ & 1.09 & $\mathrm{1.61e-}4$ & $\mathrm{3.70e-}5$ & 83.0 & 96.0 & 292 & 1330 & 1.19 & 1.24\tabularnewline
\bottomrule
\end{tabular}
\par\end{centering}
\caption{Simulation parameters and outcomes from key diagnostics. The simulations
are characterized by two input parameters: the Lundquist number $S$
and the initial noise amplitude $\epsilon$. The outcomes of diagnostics
are listed as follows: $t_{g}$ is the time when the overall fluctuation
amplitude starts to grow; $\gamma_{max,g}$ is the maximum growth
rate at $t=t_{g}$; $a_{g}$ is the current sheet half-width at $t=t_{g}$;
$t_{d}$ is the disruption time; $a_{d}$ is the current sheet half-width
and $\delta_{d}$ is the inner layer half-width at disruption; $\gamma_{d}$
is the measured growth rate and $\gamma_{max}$ is the maximum growth
rate at disruption; $k_{d}$ is the dominant wavenumber and $k_{max}$
is the fastest growing wavenumber at disruption; $t_{p}$ is the time
when the reconnection rate reaches the maximum; $t_{s}$ is the nonlinear
saturation time. These values are given in the normalized units employed
in the simulations. In the normalized units, $V_{A}=1,$ $L=0.25$,
and $\tau_{A}=0.25$.\label{tab:run_table}}
\end{table*}

\section{Diagnostics }

\label{sec_Diagnostics}
\begin{figure}
\begin{centering}
\includegraphics[scale=0.42]{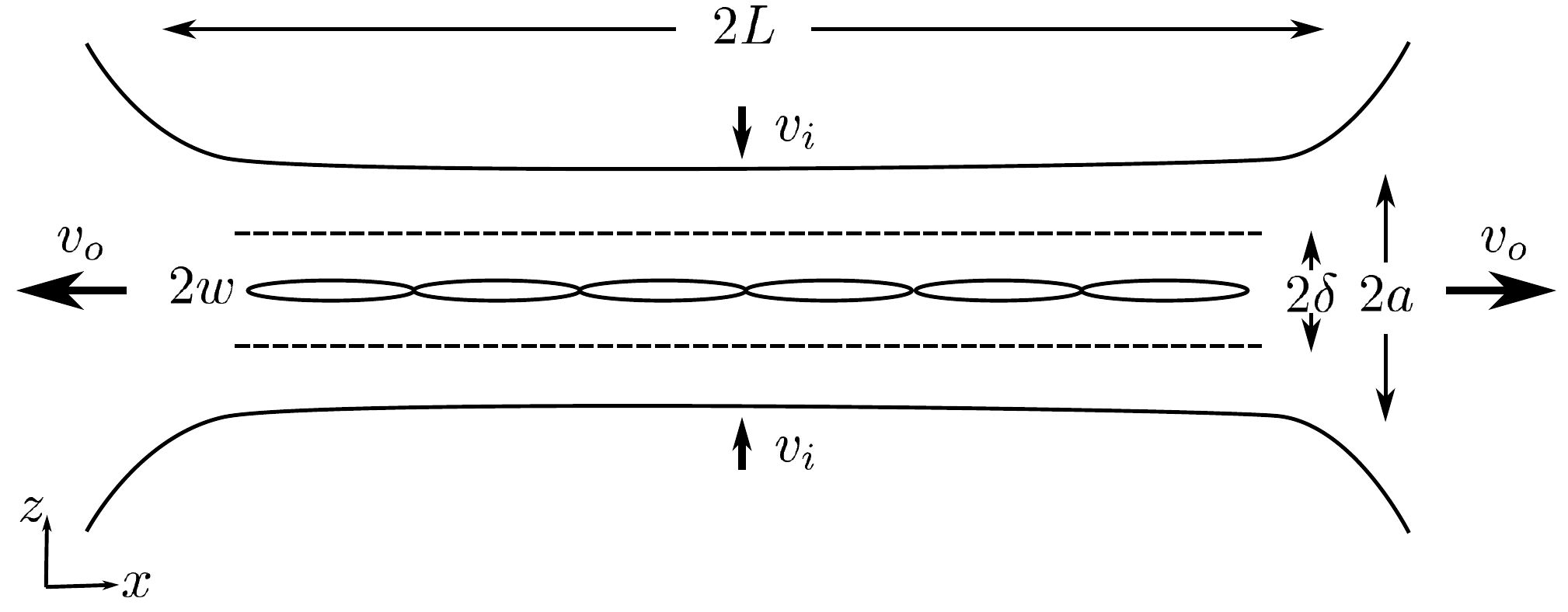}
\par\end{centering}
\caption{Schematic of plasmoid instability in a reconnecting current sheet.
Here the length of the current sheet is $2L$, and the width is $2a$.
Both the length and the width can be functions of time. The reconnection
inflows and outflows are denoted as $v_{i}$ and $v_{o}$, respectively.
Within the current sheet are two additional length scales: the inner
layer width $2\delta$, and the magnetic island width $2w$. The current
sheet is disrupted when the magnetic island width exceeds the inner
layer width. \label{fig:Schematic-of-plasmoid}}
\end{figure}
Before discussing the simulation results, we give the details of diagnostics
employed in this study. Figure \ref{fig:Schematic-of-plasmoid} shows
a schematic of plasmoid instability in a reconnecting current sheet.
Here the length of the current sheet is $2L$, and its width is $2a$.
Both the length and the width can be functions of time. The reconnection
inflows and outflows are denoted as $v_{i}$ and $v_{o}$, respectively.
There are two additional length scales within the current sheet: the
inner layer width $2\delta$ of the tearing mode, and the magnetic
island width $2w$. We call the current sheet disrupted when the magnetic
island width just exceeds the inner layer width, because at this time
the fluctuating part of the current density is approximately of the
same magnitude as the background.

In simulations, the inflows are along the $z$ direction and the outflows
are along the $x$ direction. To measure the width of the current
sheet, first we perform a tenth order polynomial fitting of $J_{y}$
along the $x$ direction between $x=-0.4$ to $x=0.4$ for each $z$
to smooth out fluctuations along $x$. Then we use the smoothed current
density profile along the $z$ direction at $x=0$ to measure the
current sheet width. In these simulations, we generally find that
the current profile is reasonably well approximated by a Harris sheet
with $J_{y}\propto\text{sech}^{2}(z/a)$. Therefore, in the following
analysis and the rest of the paper a Harris sheet profile will be
assumed. The half-width $a$ is obtained by measuring the half-width
at half-maximum (HWHM) and using the relation that the HWHM is $(\cosh^{-1}\sqrt{2})a\simeq0.882a$
for a Harris sheet. Likewise, the half-length of the current sheet
$L$ is measured using the HWHM of the current density profile along
the midplane $z=0$. The measured length $L$ is a non-monotonic function
in time, and varies within the range $0.25$ to $0.35$ before disruption
of the current sheet. As it is observed that the length $L$ does
not change substantially throughout the course of a simulation, we
assume a fixed value $L=0.25$ in the following discussion for simplicity.
Another important quantity is the gradient of outflow velocity $dv_{x}/dx$,
which determines the rate of mode-stretching along the $x$ direction.
We measure $dv_{x}/dx$ at $x=0$ by fitting a linear polynomial to
the $v_{x}$ profile along the midplane $z=0$ between $x=-0.1$ to
$x=0.1$ and then taking the derivative. 

\begin{figure}
\begin{centering}
\includegraphics[scale=0.45]{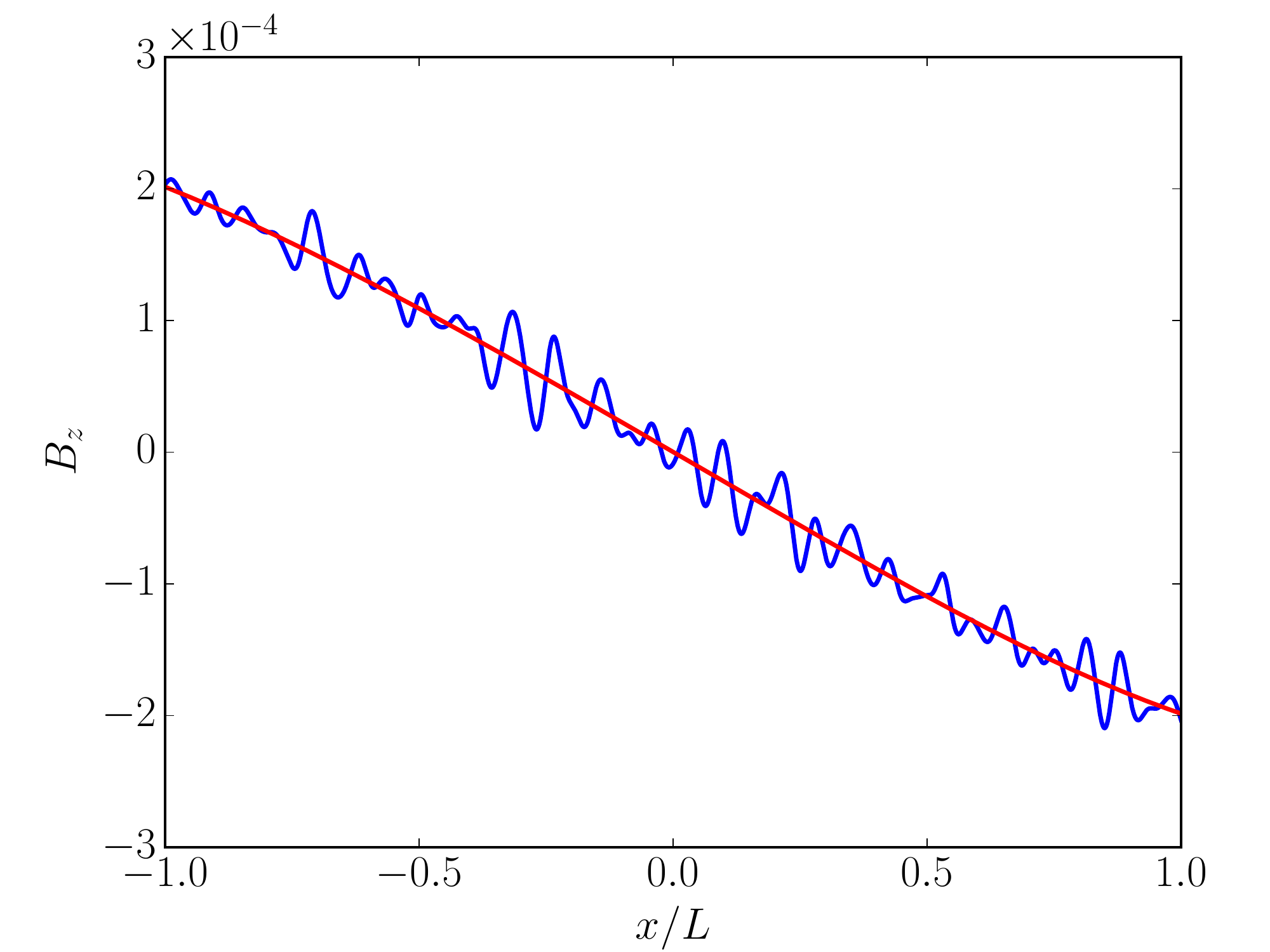}
\par\end{centering}
\caption{Separation of the fluctuation from the background. This example is
taken from Run S5 at $t=0.94$. The blue line is the total $B_{z}$
along the midplane $z=0$, and the red line is the polynomial fit
(Eq.~(\ref{eq:separation})) to determine the background. The order
of Chebyshev polynomial expansion for the fitting, $m=5$, is adaptively
determined by the procedure outlined in the text. The difference between
the total field $B_{z}$ (blue line) and the smooth background (red
line) is the fluctuation $\tilde{B}_{z}$. \label{fig:Fluctuation} }
\end{figure}

To analyze the evolution of fluctuations due to the linear instability,
fluctuations must be separated from the evolving background. Here
we employ the $B_{z}$ component of the magnetic field along the midplane
$z=0$ as the primary diagnostic. We identify the background by fitting
$B_{z}$ in $x\in[-L,L]$ as a superposition of Chebyshev polynomials
$T_{n}(x/L)$ (see\emph{, e.g.} \citet{Boyd2001}), and the remaining
part of $B_{z}$ is the fluctuation $\tilde{B}_{z}$:
\begin{equation}
B_{z}(x)=\sum_{n=0}^{m}a_{n}T_{n}(x/L)+\tilde{B}_{z}(x),\label{eq:separation}
\end{equation}
where the coefficients $a_{n}$ are determined by minimizing $\sum_{n=0}^{m}a_{n}^{2}$.
Apparently, the resulting $\tilde{B}_{z}$ from this separation depends
on the order $m$ of the polynomial expansion, especially during the
early phase of the evolution when the amplitude of the fluctuation
is much smaller than the amplitude of the background. Since the appropriate
order $m$ cannot be known \emph{a priori}, we determine $m$ adaptively
by requiring that the amplitude of the fluctuation $||\tilde{B}_{z}||\equiv\left(\int_{-L}^{L}\tilde{B}_{z}^{2}dx\right)^{1/2}$
is approximately stationary with respect to varying $m$. Specifically,
the order $m$ is determined by the following procedure. Starting
from $m=1$, we increase $m$ by $2$ in each step until the ratio
$r_{m}\equiv||\tilde{B}_{z}^{m}||/||\tilde{B}_{z}^{m-2}||$ reaches
a local maximum and $r_{m+2}\ge0.9$5. Here $\tilde{B}_{z}^{m}$ denotes
the fluctuation corresponding to a $m$-th order expansion, and we
take odd values for $m$ because $B_{z}(x)$ is approximately an odd
function. The requirement $r_{m+2}\ge0.95$ is set as a threshold
to safeguard from rare cases when the amplitude of the fluctuation
still decreases substantially with increasing $m$ immediately after
reaching a local maximum of $r_{m}$. In most cases, typically $r_{m}>0.99$
when these conditions are satisfied. An example illustrating the outcome
from this procedure is shown in Fig.~\ref{fig:Fluctuation}.

With the fluctuation $\tilde{B}_{z}$ separated from the background,
we can calculate the instantaneous growth rate as $\gamma=d\log||\tilde{B}_{z}||/dt$.
We also obtain the Fourier spectrum of $\tilde{B}_{z}$ to investigate
the evolution of dominant wave number. Because $\tilde{B}_{z}$ is
not a periodic function, we multiply $\tilde{B}_{z}$ by a $C^{\infty}$
Planck-taper window function \citep{McKechanRS2010}, which equals
unity within the range $-0.15\le x\le0.15$ and tapers off smoothly
to zero over the ranges where $0.15\le\left|x\right|\le0.25$, before
carrying out the Fourier transform. This step makes the ``windowed''
\textbf{$\tilde{B}_{z}$} vanish smoothly as $x$ approaches $\pm L$,
so that the Fourier spectrum will not be polluted by the mismatch
between $\tilde{B}_{z}(L)$ and $\tilde{B}_{z}(-L)$. Hereafter, we
denote the Fourier spectrum of the ``windowed'' \textbf{$\tilde{B}_{z}$}
as $\hat{B}_{z}(k)$, where $k$ is the wavenumber. From the amplitude
of the Fourier spectrum, we can identify the dominant wavenumber (more
on this later). Note that the growth rate calculated here is the growth
rate of the overall amplitude of the fluctuation rather than that
of the the dominant mode. However, because the Fourier spectrum is
localized around the dominant wavenumber when the plasmoid instability
is well-developed, the growth rate calculated here also approximately
represents that of the dominant mode. 

The linear instability of tearing modes is governed by the tearing
stability index $\Delta'$ \citep{FurthKR1963}, which is determined
entirely by the solution of linearized ideal MHD force-free equation
in the outer region away from the resonant surface, where $\mathbf{k}\cdot\mathbf{B}=0$.
The condition $\Delta'>0$ must be satisfied for tearing instability.
For a Harris sheet of half-width $a$, the tearing stability index
$\Delta'$ for non-oblique tearing modes is given analytically as
\begin{equation}
\Delta'=\frac{2}{a}\left(\frac{1}{ka}-ka\right),\label{eq:delta_prime}
\end{equation}
which gives the condition $ka<1$ for instability. The dispersion
relation for resistive tearing modes has been given in \citet{CoppiGPRR1976}
and requires solving a transcendental equation. Analytical approximations
can be obtained in two limits, the small-$\Delta'$ regime for short-wavelength
modes, and the large-$\Delta'$ regime for long-wavelength modes.
In the small-$\Delta'$ regime, the linear growth rate for a Harris
sheet equilibrium is given by \citep{FurthKR1963}
\begin{equation}
\gamma_{s}=C_{\Gamma}S_{a}^{-3/5}(ka)^{-2/5}(1-k^{2}a^{2})^{4/5}\frac{V_{A}}{a},\label{eq:const-psi}
\end{equation}
where $C_{\Gamma}=\left(\Gamma(1/4)/\pi\Gamma(3/4)\right)^{4/5}\simeq0.953$,
and $S_{a}\equiv aV_{A}/\eta$ is the Lundquist number defined by
the length scale $a$. In the large-$\Delta'$ regime, the linear
growth rate is approximately given by \citep{CoppiGPRR1976}
\begin{equation}
\gamma_{l}=S_{a}^{-1/3}\left(ka\right)^{2/3}\frac{V_{A}}{a}.\label{eq:nonconst-psi}
\end{equation}
The fastest growing mode takes place at the transition between the
two regimes, i.e. at $ka\sim S_{a}^{-1/4}$ when $\gamma_{s}\simeq\gamma_{l}$.
More precisely, the fastest growing wavenumber $k_{max}$ is given
by (see, \emph{e.g.} \citet{Schindler2007})

\begin{equation}
k_{max}=\frac{1.358\,S_{a}^{-1/4}}{a},\label{eq:kmax}
\end{equation}
and the corresponding growth rate is
\begin{equation}
\gamma_{max}=0.623\,S_{a}^{-1/2}\frac{V_{A}}{a}.\label{eq:gamma_max}
\end{equation}
An approximate solution that captures the two asymptotic limits (\ref{eq:const-psi})
and (\ref{eq:nonconst-psi}) is given by 
\begin{equation}
\gamma=\frac{\gamma_{s}\gamma_{l}}{\left(\gamma_{s}^{\zeta}+\gamma_{l}^{\zeta}\right)^{1/\zeta}},\label{eq:uniform_approx}
\end{equation}
for an arbitrary $\zeta$. With some experimenting we find that the
choice $\zeta=3/2$ gives a nearly exact approximation to the true
dispersion relation. 

During the linear phase of the plasmoid instability, resistivity is
important only within a narrow inner layer of thickness $2\delta$
near the resonant surface, where \citep{Biskamp1993}
\begin{align}
\delta= & \left(\dfrac{\gamma}{V_{A}/a}\frac{1}{(ka)^{2}S_{a}}\right)^{1/4}a.\label{eq:delta_inner}
\end{align}
  The plasmoid instability enters the nonlinear regime when the
magnetic island width $2w$ exceeds the inner layer width \citep{Rutherford1973}.
At this time, the electric current density fluctuation $\tilde{J}$
is of the same order of the background current density, therefore
the background current sheet loses its integrity and is ``disrupted''.
(The rationale for this description of a disrupted current sheet will
become evident in Section \ref{sec:simulation_results}.) For a single
wavelength perturbation of the form $\tilde{B}_{z}=\tilde{B}\sin(kx+\phi)$,
the island half-width is given by
\begin{equation}
w=2\sqrt{\frac{\tilde{B}}{(ka)B_{x}}}a.\label{eq:island-size}
\end{equation}
In real situations where the perturbation is a superposition of a
spectrum of modes, one has to take a combination of neighboring modes
(in the $k$ space) of the dominant mode to estimate the island size.
Here we consider a superposition over a range where $k$ varies no
more than a constant factor $\xi$, \emph{i.e.} in the range $[k/\xi,k\xi]$.
To identify the dominant wavenumber, we first smoothen the Fourier
spectrum by a running average over $[k/\xi,k\xi]$ for each $k$.
The dominant wavenumber is then determined by the peak of the smoothed
spectrum. Once the dominant wavenumber $k_{d}$ is determined, the
combined contribution of neighboring modes in the range $[k_{d}/\xi,k_{d}\xi]$
to $\tilde{B}_{z}$ can be estimated by the following relation
\begin{align}
\frac{1}{\pi}\int_{k_{d}/\xi}^{k_{d}\xi}\left|\hat{B}_{z}(k')\right|^{2}dk' & \simeq\int_{-L}^{L}\left|\tilde{B}\sin(k_{d}x)\right|^{2}dx\nonumber \\
 & \simeq\tilde{B}^{2}L,\label{eq:fluctuation}
\end{align}
which gives 
\begin{equation}
\tilde{B}=\left(\frac{1}{\pi L}\int_{k_{d}/\xi}^{k_{d}\xi}|\hat{B}_{z}(k')|^{2}dk'\right)^{1/2}.\label{eq:B_fluctuation}
\end{equation}
Using this relation for fluctuation amplitude $\tilde{B}$ in Eq.~(\ref{eq:island-size}),
we can calculate the island half-width. The disruption time $t_{d}$
is then determined as the moment when $w=\delta$ (in practice using
the first snapshot with $w\ge\delta$), where the dominant wavenumber
$k_{d}$ is used to calculate $\delta$ in Eq.~(\ref{eq:delta_inner}).
Although the procedure outlined here depends on the constant factor
$\xi$, as the spectrum $\hat{B}_{z}(k)$ is usually well-localized
at the disruption time, the result turns out to be insensitive to
the choice of $\xi$, provided that $\xi$ is not too close to unity.
We choose the value $\xi=1.5$ in this study.

\section{Simulation Results}

\label{sec:simulation_results}

\begin{figure}
\begin{centering}
\includegraphics[scale=0.7]{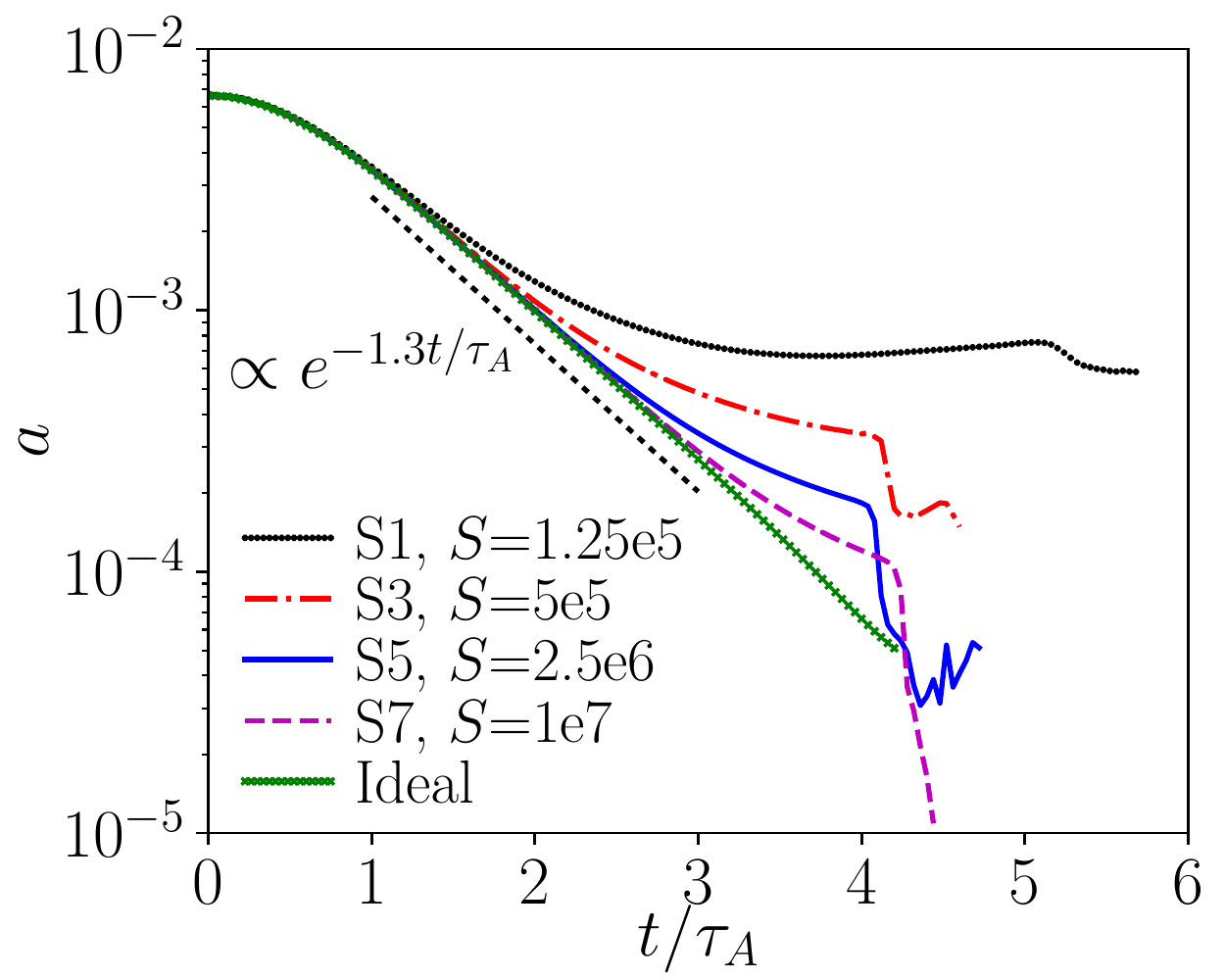}
\par\end{centering}
\caption{Time histories of the current sheet half-width $a$ for selected cases.
\textcolor{black}{Also shown for reference is the time history from
an ideal run, which exhibits a nearly perfect exponential thinning.}\textcolor{blue}{{}
}\label{fig:Time-histories-of-a}}
\end{figure}

\begin{figure*}
\begin{centering}
\includegraphics[scale=0.65]{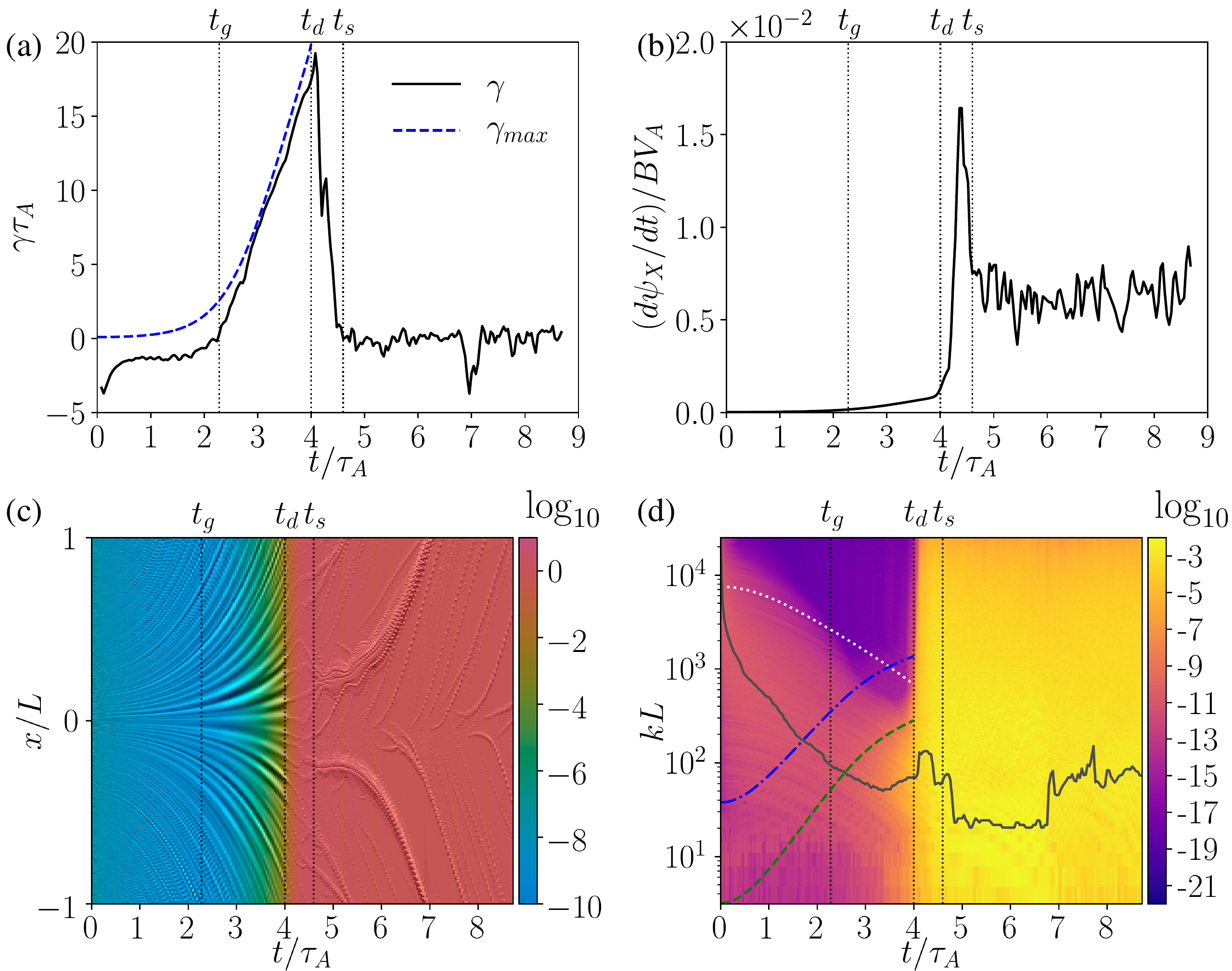}
\par\end{centering}
\caption{Diagnostics shown in time histories for Run S5. (a) The measured growth
rate $\gamma$ in solid line, and the maximum growth rate $\gamma_{max}$
in dashed line. (b) Reconnection rate. (c) Fluctuation $\tilde{B}_{z}(x)$
in real space. Here hue is used to represent the maximum fluctuation
amplitude $\max\left|\tilde{B}_{z}(x)\right|$ in logarithmic scale
at each time, and luminance is used to indicate the relative amplitude
(dark for negative and bright for positive). (d) Amplitude of the
Fourier spectrum $\left|\hat{B}_{z}(k)\right|$ in logarithmic scale.
The white dotted line indicates the trend of mode-stretching due to
the outflow jets, \emph{i.e.} $dk/dt=-kv_{x}'$ ; the blue dash-dot
line denotes the stability threshold $ka=1$; the green dashed line
denotes the fastest growing wavenumber; the gray solid line denotes
the dominant wavenumber. In all panels, $t_{g}$ marks the time when
the overall amplitude starts to grow, $t_{d}$ marks the disruption
time, and $t_{s}$ marks the nonlinear saturation time. \label{fig:Run-S5}
\textbf{}}
\end{figure*}
\begin{figure*}
\begin{centering}
\includegraphics[scale=0.55]{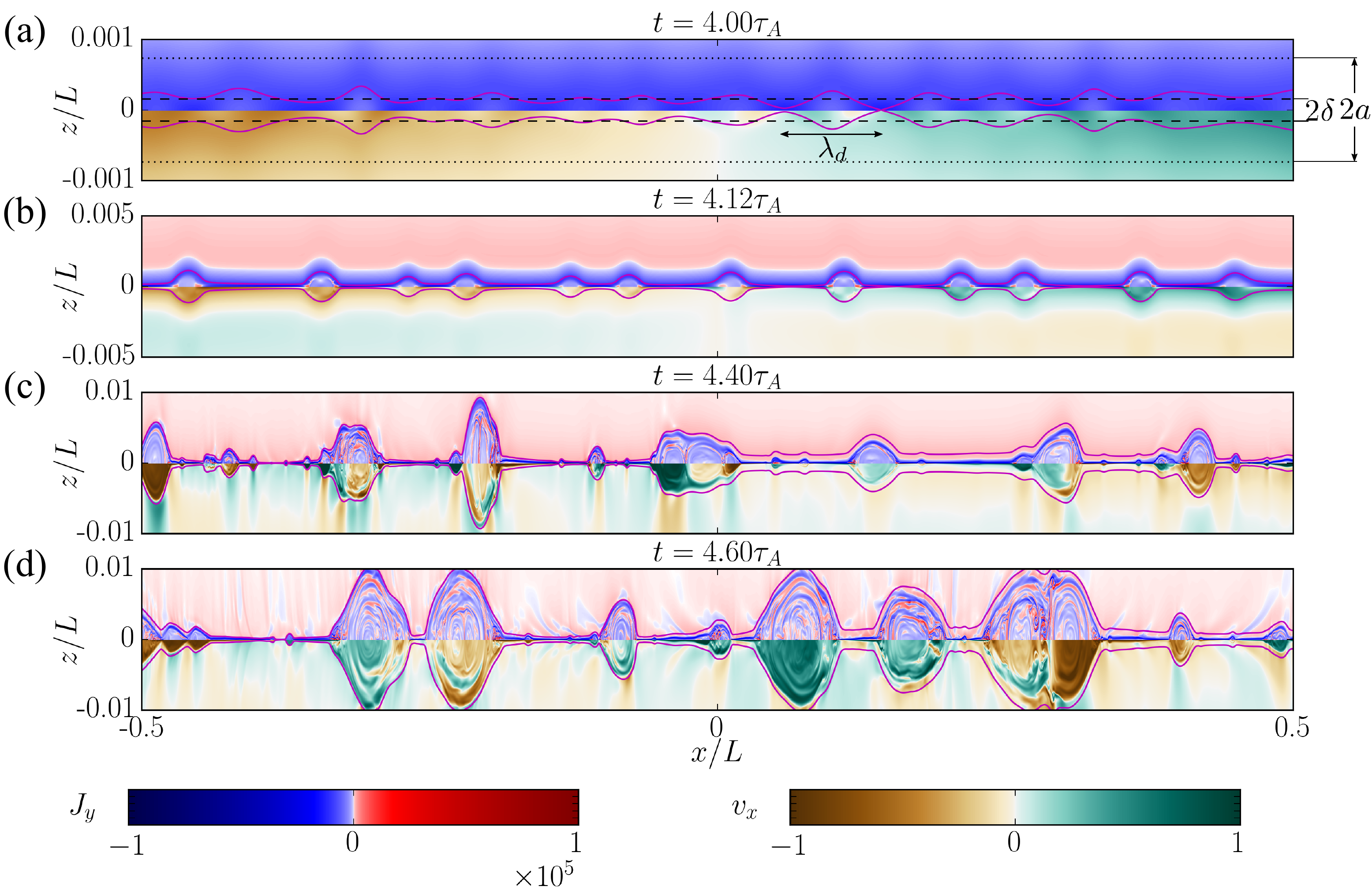}
\par\end{centering}
\caption{Representative snapshots between disruption and saturation for the
Run S5. In each panel, the upper half shows the out-of-plane current
density profile $J_{y}$ and the bottom half shows the outflow profile
$v_{x}$. The magenta solid lines denote the separatrices that separate
the two large-scale coalescing islands. Note that the $z$ direction
is stretched in these plots, and the range along $z$ direction varies
from panel to panel to better show the plasmoids. Panel (a) shows
a snapshot at the disruption time $t_{d}=4\tau_{A}$. Here the dashed
lines mark the inner layer width and the dotted lines mark the current
sheet width at disruption. The dominant wavelength identified by our
diagnostics, $\lambda_{d}=2\pi/k_{d}$, faithfully represents the
length scales of plasmoids along the $x$ direction. \textcolor{black}{Note
that the $x$ and $z$ coordinates are normalized to the current sheet
half-length $L$, as such the other length scales $\delta$, $a$,
and $\lambda_{d}$ here are also normalized in the same manner. }Panel
(b) shows a snapshot after the disruption at $t=4.12\tau_{A}$. At
this time extended secondary current sheets have developed between
plasmoids. These secondary current sheets are susceptible to the plasmoid
instability and further fragmentation. This fractal-like cascade is
clearly evident in Panel (c) at $t=4.4\tau_{A}$, when the reconnection
rate reaches the maximum. Panel (d) shows a snapshot when the plasmoid
instability reaches nonlinear saturation. \label{fig:snapshots}}
\end{figure*}

The simulations all start with an initial current sheet half-width
$a=6.66\times10^{-3}$. As time progresses, the current sheet width
decreases, as governed by the self-consistent resistive MHD equations.
The time histories of $a(t)$ for selected runs with different Lundquist
numbers $S$ are shown in Fig.~\ref{fig:Time-histories-of-a}. It
is observed that during the early period, the time history $a(t)$
is independent of $S$, indicating that the early phase of current
sheet thinning is governed by ideal MHD. After a brief initial period,
the current sheet width decreases exponentially in time according
to the relation $a(t)\propto e^{-1.3t/\tau_{A}}$, until the resistive
effect kicks in and slows down the thinning. When the plasmoid instability
disrupts the current sheet, $a(t)$ appears to exhibit a sudden plummet,
because the reconnection layer can no longer be identified as a single
current sheet, and that makes the current sheet width diagnostic inapplicable.
If the plasmoid instability does not disrupt the current sheet sufficiently
rapidly, $a(t)$ will eventually approach the Sweet-Parker width as
the asymptotic state. This is indeed the case for Run S1, where the
Lundquist number is the lowest. For other cases with higher Lundquist
numbers, the disruption takes place before the Sweet-Parker width
is approached. 

The diagnostics outlined in Sec.~\ref{sec_Diagnostics} are illustrated
in Fig.~\ref{fig:Run-S5} for Run S5. In panel (a) we show the time
history of the growth rate $\gamma$ and the maximum growth rate $\gamma_{max}$
calculated with Eq.~(\ref{eq:gamma_max}). Here (and in all other
panels) we label three important times during the evolution: (1) The
growth time $t_{g}$ marks the time when $\gamma$ goes from negative
to positive, indicating the beginning of the growth of overall fluctuation
amplitude. Here the measured growth rate is negative in the early
period because of the advection loss and dissipation in the system.
(2) The disruption time $t_{d}$ marks the disruption time as identified
by the condition $w=\delta$, and usually coincides well with the
time when $\gamma$ attains its maximum value and when $a(t)$ exhibits
a sudden plummet. After $t=t_{d}$, the growth rate $\gamma$ decreases
rapidly. (3) The saturation time $t_{s}$ marks the time beyond which
$\gamma$ is approximately zero; \emph{i.e.} the instability has reached
nonlinear saturation. In panel (b) the time history of normalized
reconnection rate is shown. Here the reconnection rate is measured
by $d\psi_{X}/dt$, where $\psi_{X}$ is the flux function $\psi$
measured at the main X-point, where the two large-scale coalescing
islands touch each other. The reconnection rate is normalized to $BV_{A}$,
which equals unity in the normalized units. It can be seen that the
reconnection rate increases rapidly after the disruption time $t_{d}$
and reaches a maximum at a time (which is denoted as $t_{p}$ in table
\ref{tab:run_table}) between $t_{d}$ and the saturation time $t_{s}$
then slows down. After the saturation time $t_{s}$, the reconnection
rate essentially remains constant and fluctuates around an average
value. In panel (c), the fluctuation $\tilde{B}_{z}(x)$ in real space
is shown. Here hue is used to represent the maximum fluctuation amplitude
$\max|\tilde{B}_{z}(x)|$ at each time in logarithmic scale, and luminance
is used to indicate the relative amplitude with respect to the overall
amplitude (dark for negative and bright for positive). The effect
of outflow jets in stretching the fluctuations is evident. Panel (d)
shows the amplitude of Fourier spectrum $|\hat{B}_{z}(k)|$ in logarithmic
scale. This panel also shows several additional lines: the white dotted
line indicates the trend of mode-stretching due to the outflow jets,
\emph{i.e.} $dk/dt=-kv_{x}'$ ; the blue dash-dot line denotes the
stability threshold $ka=1$; the green dashed line denotes the fastest
growing wavenumber; the gray solid line denotes the dominant wavenumber.
The Fourier spectrum amplitude is shown to follow the trend $dk/dt=-kv_{x}'$
closely. Another noteworthy feature is that the dominant mode at disruption
is not the fastest growing mode.  

\begin{figure*}
\begin{centering}
\includegraphics[scale=0.63]{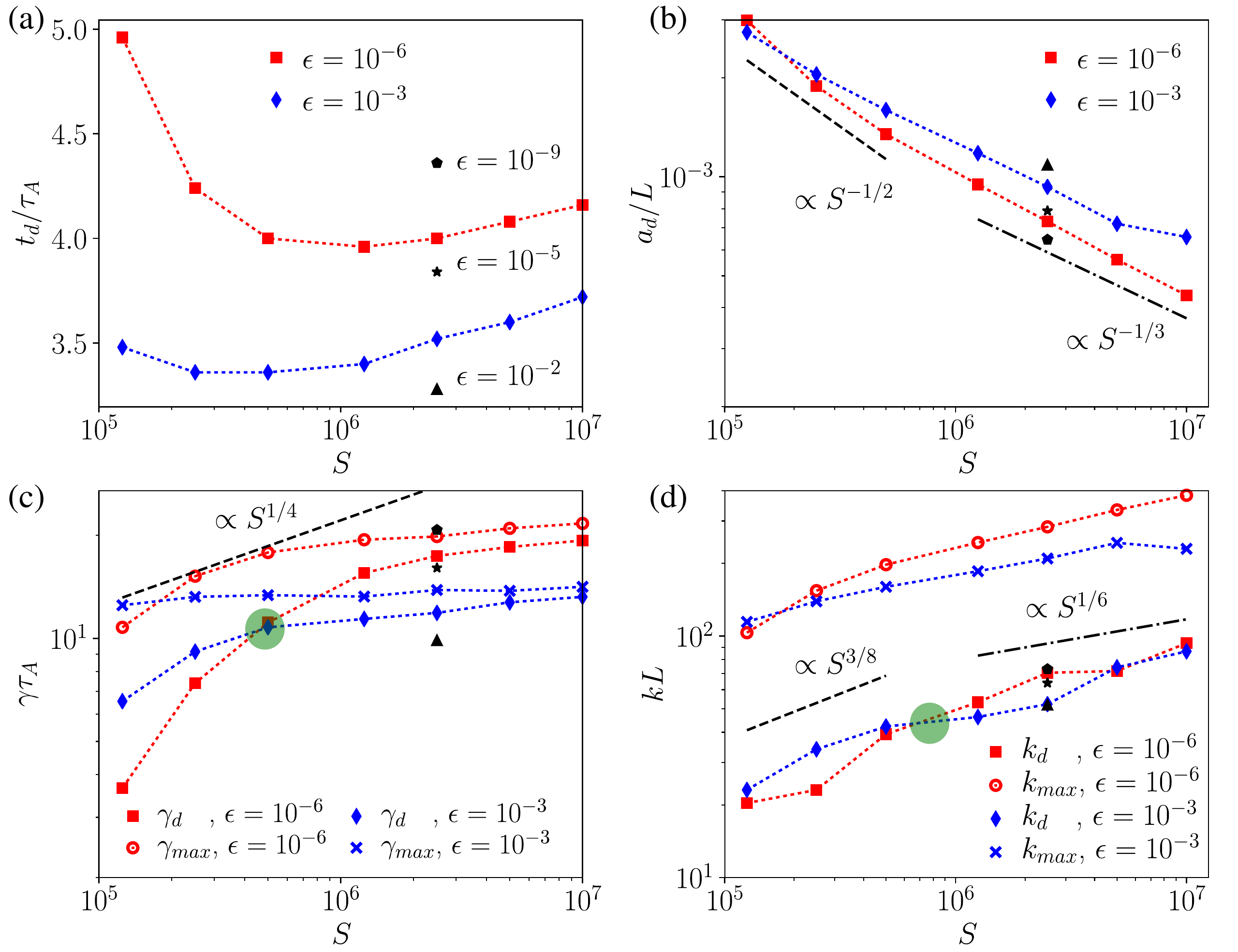}
\par\end{centering}
\caption{Scalings of the (a) disruption time $t_{d}$, (b) current sheet half-width
$a_{d}$, (c) growth rate $\gamma_{d}$, and (d) dominant wavenumber
$k_{d}$ at disruption, with respect to $S$. Here red lines correspond
to Runs S1\textendash S7, with noise amplitude $\epsilon=10^{-6}$,
and blue lines correspond to Runs H1\textendash H7, with noise amplitude
$\epsilon=10^{-3}$. Three black symbols in each panel represent Runs
N1\textendash N3 for different noise levels with $S=2.5\times10^{6}$;
see panel (a) for the legends for each symbols. In panels (c) and
(d), the growth rates and wavenumbers of the fastest mode ($\gamma_{max}$,
$k_{max}$) and the dominant mode ($\gamma_{d}$, $k_{d}$) are both
shown (only $\gamma_{d}$ and $k_{d}$ are shown for Runs N1\textendash N3).
The discrepancy between $\gamma_{max}$ and $\gamma_{d}$ are more
pronounced at low $S$ due to the effect of outflow. Note that the
dominant wavenumber is significantly smaller than that of the fastest
mode but they approximately follow the same trend.\textcolor{black}{{}
Green circles in panels (c) and (d) highlight the crossing between
two $\text{\ensuremath{\gamma}}_{d}(S)$ curves and two $k_{d}(S)$
curves with different $\epsilon$.}\label{fig:Scaling-S} }
\end{figure*}

A sequence of representative snapshots from disruption to nonlinear
saturation for Run S5 are shown in Fig.~\ref{fig:snapshots}. In
each panel, the upper half shows the out-of-plane current density
profile $J_{y}$ and the bottom half shows the outflow profile $v_{x}$.
The magenta solid lines denote the separatrices that separate the
two large-scale coalescing islands. Because the sizes of plasmoids
evolve quite substantially during this period, the range along $z$
direction varies from panel to panel to better accommodate the plasmoids
at a given time. Panel (a) shows a snapshot at the disruption time
$t_{d}=4\tau_{A}$, when typical sizes of plasmoids just exceed the
inner layer width. Here the dashed lines mark the inner layer width
and the dotted lines mark the current sheet width at disruption. The
dominant wavelength identified by our diagnostics, $\lambda_{d}=2\pi/k_{d}$,
is shown to faithfully represent the length scales of plasmoids along
the $x$ direction. In panel (b), a snapshot at $t=4.12\tau_{A}$
shows that extended secondary current sheets have developed between
plasmoids. These secondary current sheets are susceptible to the plasmoid
instability and further fragmentation. This fractal-like cascade \citep{ShibataT2001}
is clearly evident in Panel (c) at $t=4.4\tau_{A}$, when the reconnection
rate reaches the maximum. Finally, panel (d) shows a snapshot when
the plasmoid instability reaches nonlinear saturation.

Simulations of other cases all exhibit features qualitatively similar
to those presented in Figs.~\ref{fig:Run-S5} and \ref{fig:snapshots}.
However, the quantitative values of diagnostic outcomes such as the
disruption time $t_{d}$, current sheet half-width $a_{d}$, linear
growth rate $\gamma_{d}$, and dominant wavenumber $k_{d}$ depend
on the parameters $S$ and $\epsilon$ in a complex manner. The outcomes
of diagnostics are summarized in Table~\ref{tab:run_table}, and
plotted in Fig.~\ref{fig:Scaling-S} as scalings with respect to
the Lundquist number $S$. The values in Table~\ref{tab:run_table}
are given in the original normalized units of the simulations. On
the other hand, in Fig.~\ref{fig:Scaling-S}, the quantities $t_{d}$,
$a_{d}$, $\gamma_{d}$, and $k_{d}$ are normalized to appropriate
global scales $\tau_{A}$ and $L$; recall that in the original normalized
units, $V_{A}=1,$ $L=0.25$, and $\tau_{A}=0.25$. The panels in
Fig.~\ref{fig:Scaling-S} are organized in two sets of parameter
scans with respect to $S$. Runs S1\textendash S7 with noise amplitude
$\epsilon=10^{-6}$ are shown in red symbols, and runs H1\textendash H7
with $\epsilon=10^{-3}$ are shown in blue. Three additional runs
N1\textendash N3 with varying noise amplitudes at the same $S=2.5\times10^{6}$
are shown in black symbols.

The growth time $t_{g}$ marks the beginning when the overall amplitude
starts to grow. It can be seen from Table~\ref{tab:run_table} that
the maximum growth rate $\gamma_{max,g}=O(10)$, \emph{i.e.}~$\gamma_{max,g}\tau_{A}\simeq2\,\text{--}\,3$,
at $t=t_{g}$ for all cases. This reaffirms the physical intuition
that for the fluctuation amplitude to grow effectively, the growth
rate has to be greater than the inverse of the time scale for current
sheet evolution as well as that for the ejection of outflow jets,
both of the order of $\tau_{A}$. As will be made clear in Section
\ref{sec:Phenomenological-Model}, the overall amplitude of fluctuations
can grow only when the linear growth from tearing modes is sufficient
to overcome the decrease caused by advection loss and mode-stretching
due to the outflow. Because at a higher $S$, the current sheet must
thin down to a smaller width to achieve a maximum growth rate with
$\gamma_{max,g}\tau_{A}\simeq2\,\text{--}\,3$, it is observed that
$t_{g}$ increases monotonically with $S.$ On the other hand, $t_{g}$
is mostly independent of the initial noise amplitude $\epsilon$,
as long as the initial perturbation is sufficiently small to remain
in the linear regime.

The disruption time $t_{d}$, shown in Fig.~\ref{fig:Scaling-S}(a),
is found to be non-monotonic with respect to $S$. This can be understood
qualitatively as follows. At low Lundquist numbers, the current sheet
width approaches the Sweet-Parker width before disruption. Since the
plasmoid instability growth rate scales as $S^{1/4}$ for Sweet-Parker
current sheets, it takes longer to disrupt a low-$S$ current sheet
because the growth rate is lower. On the other hand, at high Lundquist
numbers, the current sheet will be disrupted before it approaches
the Sweet-Parker width. However, the maximum growth rate has to reach
$\gamma_{max}\tau_{A}=O(1)$ for the fluctuation to start growing,
and it takes longer to reach that point at higher $S$. Consequently,
the disruption time also becomes longer.\textbf{ } 

Figure \ref{fig:Scaling-S}(b) shows the scalings of current sheet
half-width $a_{d}$ at disruption. The scaling for cases with $\epsilon=10^{-6}$
(red symbols) clearly shows a change in the scaling behavior as $S$
increases. At low $S$, the scaling is close to $a_{d}\sim S^{-1/2}$,
indicating that the current sheet is approaching the Sweet-Parker
width at disruption. At high $S$, the scaling becomes less steep
and is close to $a_{d}\propto S^{-1/3}$. For cases with $\epsilon=10^{-3}$
(blue symbols), the scaling is close to $a_{d}\propto S^{-1/3}$ over
the entire range of $S$ we have scanned. Generally speaking, for
fixed Lundquist number, the time to disruption is longer when the
initial noise is smaller, and therefore, $a_{d}$ is smaller. This
trend is evident for the runs S5, H5, and N1\textendash N3 with $S=2.5\times10^{6}$.
However, we note an exception to this rule for runs S1 and H1, both
at $S=1.25\times10^{5}$, where $a_{d}$ is found to be larger for
run S1, even though the initial noise is smaller. This happens because
$a(t)$ is not necessarily a monotonically decreasing function. It
is well-known that the two large-scale merging islands can bounce
back, in a phenomenon known as ``sloshing'', after the distance
between them reaches a minimum \citep{KnollC2006}. When the two islands
bounce back, the current sheet half-width $a(t)$ increases. It can
be seen from Fig.~\ref{fig:Time-histories-of-a} that disruption
occurs when $a(t)$ is weakly increasing for run S1, therefore $a_{d}$
for run S1 is larger than that for run H1, in which disruption occurs
at an earlier time. Note that sloshing takes place before disruption
only at low $S$, because at high $S$, disruption usually takes place
before the distance between the two large-scale islands reaches a
minimum. 

The scalings of the linear growth rate at disruption are shown in
Fig.~\ref{fig:Scaling-S}(c). Here the scalings for two growth rates
are shown: the growth rate $\gamma_{d}$ measured from simulations,
and the fastest growth rate $\gamma_{max}$ calculated from Eq.~(\ref{eq:gamma_max})
with $a=a_{d}$. The measured growth rate $\gamma_{d}$ is smaller
than the maximum growth rate $\gamma_{max}$, which is understandable
as the dominant mode is not the fastest growing mode. The discrepancy
between $\gamma_{d}$ and $\gamma_{max}$ is more pronounced at lower
$S$, because mode-stretching due to outflow jets becomes more effective
(more on this in Sec.~\ref{sec:Phenomenological-Model}). For cases
with $\epsilon=10^{-6}$, the maximum growth rate $\gamma_{max}\sim S^{1/4}$
at lower $S$ and becomes nearly independent of $S$ at higher $S$;
for cases with $\epsilon=10^{-3}$, $\gamma_{max}$ is approximately
independent of $S$ in the range we have scanned. These scalings are
direct consequences of the scalings of $a_{d}$ we discussed previously.
Likewise, the scalings of the dominant wavenumber $k_{d}$ and the
fastest growing wavenumber $k_{max}$ at disruption are shown in Fig.~\ref{fig:Scaling-S}(d).
Here, the dominant wavenumber $k_{d}$ is observed to be smaller than
the fastest growing wavenumber $k_{max}$ by a factor of approximately
3 to 6, where the factor tends to be larger at lower $S$. Because
the spectrum $\hat{B_{z}}(k)$ at disruption is not a very sharp peak,
there is considerable uncertainty in determining the dominant wavenumber,
and this is reflected in the irregular appearance of the scalings
of $k_{d}$. We should point out that the various power-law scalings
shown in Fig.~\ref{fig:Scaling-S} (b)\textendash (d) are only for
reference; they should not be interpreted as the actual scalings of
those quantities. 

A notable feature in Fig.~\ref{fig:Scaling-S}(c) is the crossing,
highlighted by the green circle, between the two $\gamma_{d}(S)$
curves for different initial noise amplitudes. Generally speaking,
if the noise amplitude is lower, it takes longer to disrupt the current
sheet, hence the current sheet is thinner at disruption and the growth
rate is higher. This trend is observed at high $S$ but the trend
is the opposite at low $S$, leading to the crossing of the two $\gamma_{d}(S)$
curves. A similar crossing is also observed for the two $k_{d}(S)$
curves in Fig.~\ref{fig:Scaling-S}(d), although it is less clean
due to the greater uncertainty in obtaining $k_{d}$. As will be discussed
in greater detail in Sec.~\ref{sec:Phenomenological-Model}, this
``crossing'' between $S$-dependence curves is due to the stretching
effect of the outflow, which causes a significant departure of the
dominant mode from the fastest growing mode at low $S$. Ultimately,
this crossing is linked to the notion of the ``critical'' Lundquist
number $S_{c}$ for the plasmoid instability, which turns out to depend
on the initial noise amplitude instead of being a constant value.

\section{Phenomenological Model For Current Sheet Disruption}

\label{sec:Phenomenological-Model}

The simulation results presented in Sec.~\ref{sec:simulation_results}
regarding the scalings of various quantities at disruption can be
reproduced, at least qualitatively, with a simple model. In this model,
we seek a description for the time evolution of fluctuation in the
Fourier space, in terms of a normalized amplitude $f(k,t)\equiv|\hat{B}_{z}(k,t)|/B_{0}L_{0},$
where $B_{0}$ is a characteristic magnetic field and $L_{0}$ is
a characteristic length. In this model, we allow the current sheet
length and width to evolve in time. 

Assuming that the outflow velocity $v_{x}$ profile is approximately
linear, the mode-stretching due to outflow implies that $f(k,t)$
will be advected in $k$-space following the time evolution of wavenumber
$k$:
\begin{equation}
\frac{dk}{dt}=-v_{x}'k,\label{eq:mode-stretching}
\end{equation}
where $v_{x}'=dv_{x}/dx$ is the gradient of the outflow velocity.
However, the fluctuation amplitude $f(k,t)$ can also change when
it is advected, as will be determined as follows. 

Taking the $z$ component of the linearized ideal induction equation
along the mid-plane $z=0$ yields 
\begin{equation}
\partial_{t}\tilde{B}_{z}+v_{x}\tilde{B}_{z}'=-\tilde{B}_{z}v_{x}'.\label{eq:induction}
\end{equation}
Here the left-hand-side of Eq.~(\ref{eq:induction}) represents\textbf{
}the rate of change of $\tilde{B}_{z}$ in the reference frame moving
with the plasma, which is equal to the right-hand-side representing
the change of $\tilde{B}_{z}$ due to stretching or compression. Because
we are interested in fluctuations within the current sheet, we seek
the time dependence of the $L^{2}$-norm of $\tilde{B}_{z}$ over
the current sheet region\emph{ $\left|\left|\tilde{B}_{z}\right|\right|^{2}\equiv\int_{-L}^{L}\tilde{B}_{z}^{2}dx$.}
First, we multiply Eq.~(\ref{eq:induction}) by $\tilde{B}_{z}$
and integrate from $x=-L$ to $L$ to yield (after integrating by
parts and multiplying by a factor of two)
\begin{equation}
\int_{-L}^{L}\partial_{t}\tilde{B}_{z}^{2}dx=-\left.v_{x}\tilde{B}_{z}^{2}\right|_{-L}^{L}-v_{x}'\int_{-L}^{L}\tilde{B}_{z}^{2}dx,\label{eq:integral_induction}
\end{equation}
where the first term on the right hand side accounts for the advection
loss, and the second term is the contribution from stretching. To
evaluate the first term we need $v_{x}$ and $\tilde{B}_{z}^{2}$
at the end points $x=\pm L$. The outflow velocity $v_{x}=\pm v_{o}$
at $x=\pm L$, but $\tilde{B}_{z}^{2}$ at end points fluctuate in
time and cannot been known beforehand. To proceed, we further assume
that the amplitude of $\tilde{B}_{z}$ is approximately uniform along
the current sheet, and replace $\tilde{B}_{z}^{2}$ at $x=\pm L$
with the averaged value $\frac{1}{2L}\int_{-L}^{L}\tilde{B}_{z}^{2}dx$
over the entire current sheet length. This approximation is valid
if the time scale of interest (\emph{i.e.} the time scale on which
$\left|\left|\tilde{B}_{z}\right|\right|^{2}$ evolves) is longer
than the typical fluctuating time scale of $\tilde{B}_{z}$ at the
end points, which is justified according to our simulations (see Fig.~\ref{fig:Run-S5}(c)).
Under this assumption, we obtain the advection loss term $\left.v_{x}\tilde{B}_{z}^{2}\right|_{-L}^{L}\simeq\frac{v_{o}}{L}\int_{-L}^{L}\tilde{B}_{z}^{2}dx=v_{x}'\int_{-L}^{L}\tilde{B}_{z}^{2}dx$,
which is the same as the contribution from stretching. Hence, Eq.~(\ref{eq:integral_induction})
can be written as
\begin{equation}
\int_{-L}^{L}\partial_{t}\tilde{B}_{z}^{2}dx=-2v_{x}'\int_{-L}^{L}\tilde{B}_{z}^{2}dx.\label{eq:energy_decay}
\end{equation}
Taking into account that the current sheet length $L$ can evolve
in time and applying Leibniz's rule, we obtain the time rate of change
of the $L^{2}$-norm of $\tilde{B}_{z}$ over the current sheet region
\begin{align}
\frac{d}{dt}\int_{-L}^{L}\tilde{B}_{z}^{2}dx & =\int_{-L}^{L}\partial_{t}\tilde{B}_{z}^{2}dx+\frac{dL}{dt}\left(\left.\tilde{B}_{z}^{2}\right|_{x=L}+\left.\tilde{B}_{z}^{2}\right|_{x=-L}\right)\nonumber \\
 & =\left(-2v_{x}'+\frac{1}{L}\frac{dL}{dt}\right)\int_{-L}^{L}\tilde{B}_{z}^{2}dx.\label{eq:integral_induction1}
\end{align}
Here, in the second step we apply Eq. (\ref{eq:energy_decay}) for
the first term on the right hand side, and again approximate $\tilde{B}_{z}^{2}$
at $x=\pm L$ with the averaged value $\frac{1}{2L}\int_{-L}^{L}\tilde{B}_{z}^{2}dx$.
Equation (\ref{eq:integral_induction1}) accounts for the change of
the $L^{2}$-norm due to the combined effects of stretching, advection
loss, and evolution of the current sheet length, where the former
two effects give the same contribution (hence the factor of two in
the $v_{x}'$ term). For the special case that the current sheet is
lengthening with the same speed as the outflow, \emph{i.e.} $dL/dt=v_{o}$,
the latter two effects exactly cancel each other because the outflow
cannot escape from the current sheet, therefore only the contribution
from stretching effect remains.

The next step is to rewrite the $L^{2}$-norm of $\tilde{B}_{z}$
in Fourier space using the relation $\int_{0}^{\infty}|\hat{B}_{z}|^{2}dk=\pi\int_{-L}^{L}|\tilde{B}_{z}|^{2}dx$
in Eq.~(\ref{eq:integral_induction1}). The time derivative of $\int_{0}^{\infty}|\hat{B}_{z}|^{2}dk$
can be expressed as
\begin{align}
\frac{d}{dt}\int_{0}^{\infty}|\hat{B}_{z}|^{2}dk & =\int_{0}^{\infty}\frac{d}{dt}\left(|\hat{B}_{z}|^{2}dk\right)\nonumber \\
 & =\int_{0}^{\infty}\frac{d|\hat{B}_{z}|^{2}}{dt}dk+\int_{0}^{\infty}|\hat{B}_{z}|^{2}\frac{d}{dt}dk\nonumber \\
 & =\int_{0}^{\infty}\frac{d|\hat{B}_{z}|^{2}}{dt}dk-\int_{0}^{\infty}|\hat{B}_{z}|^{2}v_{x}'dk,\label{eq:t-derivative}
\end{align}
where the time derivative $d/dt$ on the right hand side is the Lagrangian
time derivative in Fourier space following the stretching of wavenumber
$k$ in Eq.~(\ref{eq:mode-stretching}), \emph{i.e.
\begin{equation}
\frac{d}{dt}=\frac{\partial}{\partial t}+\frac{dk}{dt}\frac{\partial}{\partial k}=\frac{\partial}{\partial t}-kv_{x}'\frac{\partial}{\partial k}.\label{eq:ddt}
\end{equation}
}Note that because the wavenumber $k$ evolves in time from the Lagrangian
viewpoint, the differential wavenumber $dk$ also evolves according
to the relation $d(dk)/dt=-v_{x}'dk$. Alternatively, Eq.~(\ref{eq:t-derivative})
can be derived from the Eulerian viewpoint and applying Eq.~(\ref{eq:ddt}):
\begin{align}
\frac{d}{dt}\int_{0}^{\infty}|\hat{B}_{z}|^{2}dk= & \int_{0}^{\infty}\frac{\partial}{\partial t}|\hat{B}_{z}|^{2}dk\nonumber \\
= & \int_{0}^{\infty}\frac{d}{dt}|\hat{B}_{z}|^{2}dk+\int_{0}^{\infty}kv_{x}'\frac{\partial}{\partial k}|\hat{B}_{z}|^{2}dk\nonumber \\
= & \int_{0}^{\infty}\frac{d}{dt}|\hat{B}_{z}|^{2}dk-\int_{0}^{\infty}|\hat{B}_{z}|^{2}v_{x}'dk\nonumber \\
 & +\left.kv_{x}'|\hat{B}_{z}|^{2}\right|_{k=0}^{k=\infty}.\label{eq:ddt1}
\end{align}
Assuming that $|\hat{B}_{z}|^{2}$ vanishes at $k=0$ and $\infty$,
we obtain the same relation as Eq.~(\ref{eq:t-derivative}). 

Using Eq.~(\ref{eq:t-derivative}) in Eq. (\ref{eq:integral_induction1})
yields 
\begin{equation}
\int_{0}^{\infty}\frac{d|\hat{B}_{z}|^{2}}{dt}dk=\left(-v_{x}'+\frac{1}{L}\frac{dL}{dt}\right)\int_{0}^{\infty}|\hat{B}_{z}|^{2}dk.\label{eq:t-derivative-k-space}
\end{equation}
Here the right hand side represents the effects of advection loss
and evolution of the current sheet length, while the stretching effect
is incorporated in the Lagrangian time derivative on the left hand
side. From Eq.~(\ref{eq:t-derivative-k-space}), we obtain the Lagrangian
time derivative of the Fourier component amplitude:\footnote{The reader may wonder how the Fourier amplitude can change due to
change in $L$. The reason is that we are considering fluctuations
within the current sheet, from $x=-L$ to $x=L$, and project that
onto the Fourier components. Hence the projection becomes larger (smaller)
when $L$ increases (decreases). } 
\begin{equation}
\frac{d|\hat{B}_{z}|}{dt}=\left(-\frac{v_{x}'}{2}+\frac{1}{2L}\frac{dL}{dt}\right)|\hat{B}_{z}|.\label{eq:rate_change}
\end{equation}
As a consistency check, note that this equation also gives the correct
physical condition that magnetic island width does not change due
to stretching, as can be verified by taking the time derivative of
the island half-width $w$ from Eq.~(\ref{eq:island-size}), applying
Eq.~(\ref{eq:B_fluctuation}) and Eq.~(\ref{eq:rate_change}).

The model is completed by adding the contribution from the linear
growth of tearing instability $\gamma|\hat{B}_{z}|$ to the right
hand side of Eq.~~(\ref{eq:rate_change}), which gives the time
evolution equation for the fluctuation amplitude $|\hat{B}_{z}|$
in Fourier space. The final model equation is expressed in terms of
the normalized fluctuation amplitude, $f(k,t)\equiv|\hat{B}_{z}(k,t)|/B_{0}L_{0}$,
as: 
\begin{equation}
\frac{df}{dt}=\partial_{t}f-kv_{x}'\partial_{k}f=\left(\gamma-\frac{v_{x}'}{2}+\frac{1}{2L}\frac{dL}{dt}\right)f.\label{eq:linear-growth}
\end{equation}
Here, the growth rate $\gamma$ depends on time through the time dependence
of the current sheet half-width $a(t)$ and the upstream magnetic
field $B(t)$. We further limit the domain to the region $k\ge k_{min}\equiv\pi/L$,
below which the wavelength cannot be contained in the current sheet.
If we take the Lagrangian viewpoint of following the stretching of
a mode according to Eq.~(\ref{eq:mode-stretching}), the wavelength
of the mode becomes longer over time. At a certain point, the wavelength
will be stretched to become longer than the length of the current
sheet, and we can say that the mode is advected out of the current
sheet.

It can be shown from Eq.~(\ref{eq:linear-growth}) that the time
rate of change of the $L^{2}$-norm of the fluctuation, limited to
the domain $k\ge k_{min}$, is given by
\begin{multline}
\frac{d}{dt}\int_{k_{min}}^{\infty}f^{2}dk=\int_{k_{min}}^{\infty}2\gamma f^{2}\,dk-v_{x}'\int_{k_{min}}^{\infty}f^{2}dk\\
+\left(\frac{1}{L}\frac{dL}{dt}-v_{x}'\right)\left(\left.kf^{2}\right|_{k_{min}}+\int_{k_{min}}^{\infty}f^{2}dk\right).\label{eq:d_norm_dt}
\end{multline}
The first term on the right hand side of Eq.~(\ref{eq:d_norm_dt})
accounts for the increase in fluctuation amplitude due to tearing
modes. The second term is the decrease due to stretching effect. The
third term accounts for the effects of advection loss and the evolution
of current sheet length, where the surface term $\left.kf^{2}\right|_{k_{min}}$
comes from integration by parts and application of Leibniz's rule
on the lower bound $k_{min}$ of the integral (note that $dk_{min}/dt=-(k_{min}/L)dL/dt$.)
The effects of advection loss and the evolution of current sheet length
exactly cancel each other when the current sheet is lengthening with
the outflow speed, \emph{i.e.} $dL/dt=v_{o}$, as the outflow cannot
escape from the current sheet. However, in general we should expect
$dL/dt\le v_{o}$, hence the third term on the right hand side of
Eq.~(\ref{eq:d_norm_dt}) is also negative. As such, to have a positive
growth of the $L^{2}$-norm, the linear growth rate of tearing mode
has to be sufficiently large to overcome the decrease due to advection
loss and stretching. Under most circumstances, the surface term $\left.kf^{2}\right|_{k_{min}}$
is negligible compared to $\int_{k_{min}}^{\infty}f^{2}dk$, as long
as the dominant wavenumber is much larger than $k_{min}$. If we further
assume that $|dL/dt|\ll v_{o}$, then 
\begin{equation}
\frac{d}{dt}\int_{k_{min}}^{\infty}f^{2}dk\simeq\int_{k_{min}}^{\infty}2(\gamma-1/\tau_{A})f^{2}\,dk.\label{eq:d_norm_dt1}
\end{equation}
Since $\gamma$ is a function of $k$, Eq.~(\ref{eq:d_norm_dt1})
gives $\gamma_{max}\tau_{A}>1$ as a minimum requirement for a positive
growth of the overall fluctuation amplitude, which is consistent with
our simulation findings. 

Equation (\ref{eq:d_norm_dt}) implies that if the plasmoid instability
does not exist, the fluctuation amplitude will decrease monotonically
in time and asymptotically approach zero as $t\to\infty$.  However,
since noise is always present in any natural systems, noise will replenish
the fluctuation and prevent it from approaching zero even in the absence
of the plasmoid instability. Noise can be introduced into the model
by explicitly adding a source term or by setting a floor to the fluctuation
amplitude as a lower bound. To properly model noise in the system
requires extra knowledge of the environment, and will not be discussed
further here. For the cases we consider here, the plasmoid instability
sets in sufficiently rapidly and the decay of fluctuation amplitude
is not a significant issue. 

Up to this point, the time dependences of the current sheet half-width
$a$, the half-length $L$, the upstream magnetic field $B$, and
the velocity gradient $v_{x}'$ are not specified. These conditions
must be provided according to the physical system under consideration.
To fix the idea, we consider a situation when $L$ and $B$ are no
longer evolving, \emph{i.e.} $L=L_{0}$ and $B=B_{0}$. Let $v_{i}$
and $v_{o}$ be the inflow and the outflow speeds, respectively. Assuming
an incompressible plasma, conservation of mass and energy implies
that $v_{o}=B_{0}/\sqrt{\rho}=V_{A}$ and $v_{i}=aV_{A}/L_{0}$, as
in the Sweet-Parker model. Therefore, the velocity gradient $v_{x}'=V_{A}/L=1/\tau_{A}$
is independent of time. The time-dependence of the half-width $a(t)$
can be obtained by integrating the induction equation $\partial_{t}\mathbf{B}=-\nabla\times\mathbf{E}$
from $z=0$ to the asymptotic outer region $z=z_{out}\gg a$, which
yields
\begin{equation}
\frac{d}{dt}\int_{0}^{z_{out}}B_{x}dz=\left.E_{y}\right|_{z_{out}}-\left.E_{y}\right|_{0}.\label{eq:induction_integrate}
\end{equation}
Assuming a Harris sheet profile $B_{x}=B_{0}\tanh(z/a)$ and formally
setting $z_{out}\to\infty$, using the relations $\left.E_{y}\right|_{\infty}=v_{i}B_{0}=aV_{A}B_{0}/L_{0}$
and $\left.E_{y}\right|_{0}=\left.\eta J_{y}\right|_{z=0}=\eta B_{0}/a$,
we obtain
\begin{equation}
\frac{d}{dt}\int_{0}^{\infty}B_{0}\tanh(z/a)dz=\frac{aV_{A}}{L}B_{0}-\eta\frac{B_{0}}{a}.\label{eq:induction1}
\end{equation}
The left-hand-side can be calculated using
\begin{align}
\frac{d}{dt}\int_{0}^{\infty}\tanh(z/a)dz & =-\frac{da}{dt}\int_{0}^{\infty}\frac{z}{a^{2}}\mbox{sech}^{2}(z/a)dz\nonumber \\
 & =-\log(2)\frac{da}{dt},\label{eq:LHS}
\end{align}
hence we obtain a time evolution equation for $a(t)$:
\begin{equation}
\log(2)\frac{da}{dt}+\frac{V_{A}}{L}a-\frac{\eta}{a}=0.\label{eq:a-evolution}
\end{equation}
The solution is 
\begin{equation}
a^{2}=a_{SP}^{2}+(a_{0}^{2}-a_{SP}^{2})\exp(-(2/\log2)(t/\tau_{A})),\label{eq:a-solution}
\end{equation}
where $a_{0}$ is the half-width at $t=0$ and $a_{SP}=L/\sqrt{S}$
is the Sweet-Parker width.\footnote{An alternative derivation of a similar functional form can be found
in \citet{Kulsrud2005}, page 425 \textendash{} 426.} This solution approaches $a_{SP}$ in the asymptotic limit $t\to\infty$.
At high $S$ with $a_{SP}\ll a_{0}$, the half-width decays exponentially
with $a\simeq a_{0}\exp(-1.44\,t/\tau_{A})$ at early time. This is
in good agreement with what we obtain in simulations, shown in Fig.~\ref{fig:Time-histories-of-a},
apart from an initial transient period before the Alfv\'enic outflow
jets have been established.

\begin{figure*}
\begin{centering}
\includegraphics[scale=0.65]{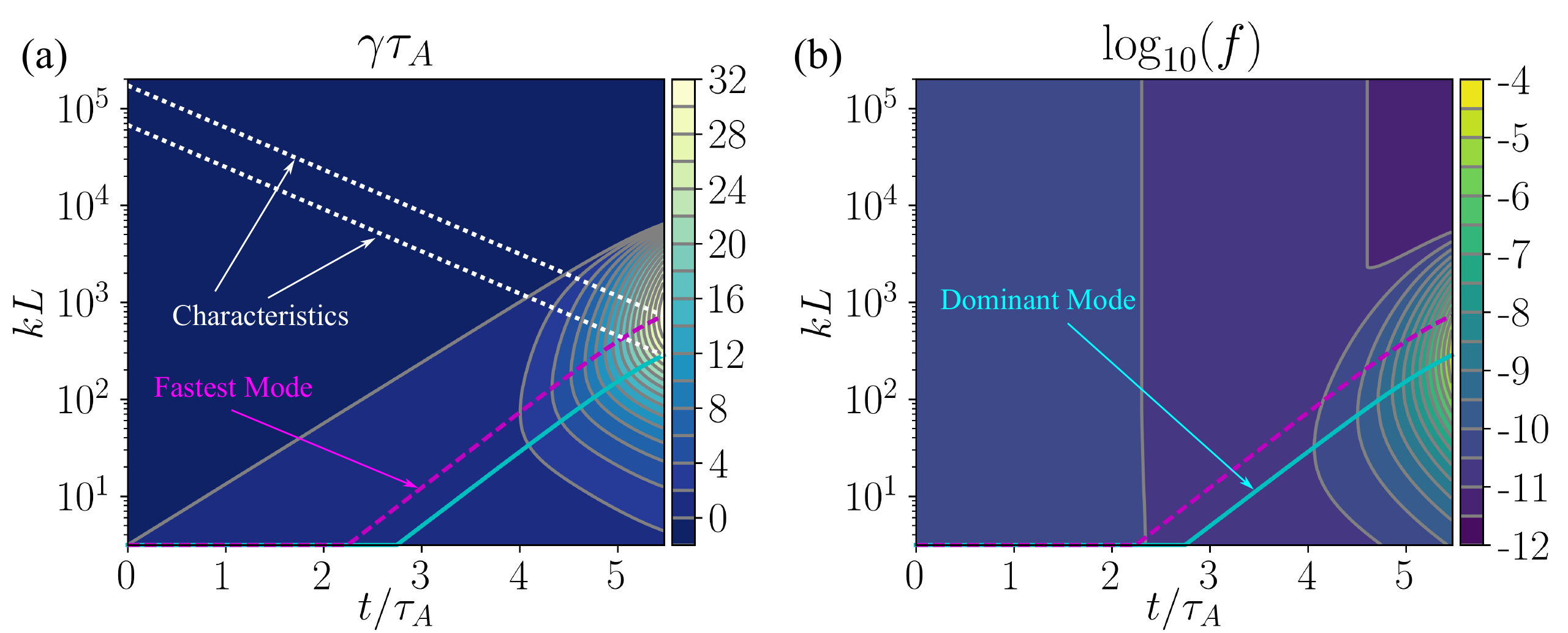}
\par\end{centering}
\caption{(a) Linear growth rate, and (b) solution of the model equation (\ref{eq:linear-growth})
for the case $S=10^{8}$, $\epsilon=10^{-10}$. Also marked in the
panels are the wavenumbers for the fastest growing mode (magenta dashed
line) and the dominant mode (cyan solid line). The white dotted lines
in panel (a) are characteristics corresponding to the dominant mode
and the fastest mode at disruption. \label{fig:model-solution}}
\end{figure*}
\begin{figure*}
\begin{centering}
\includegraphics[scale=0.63]{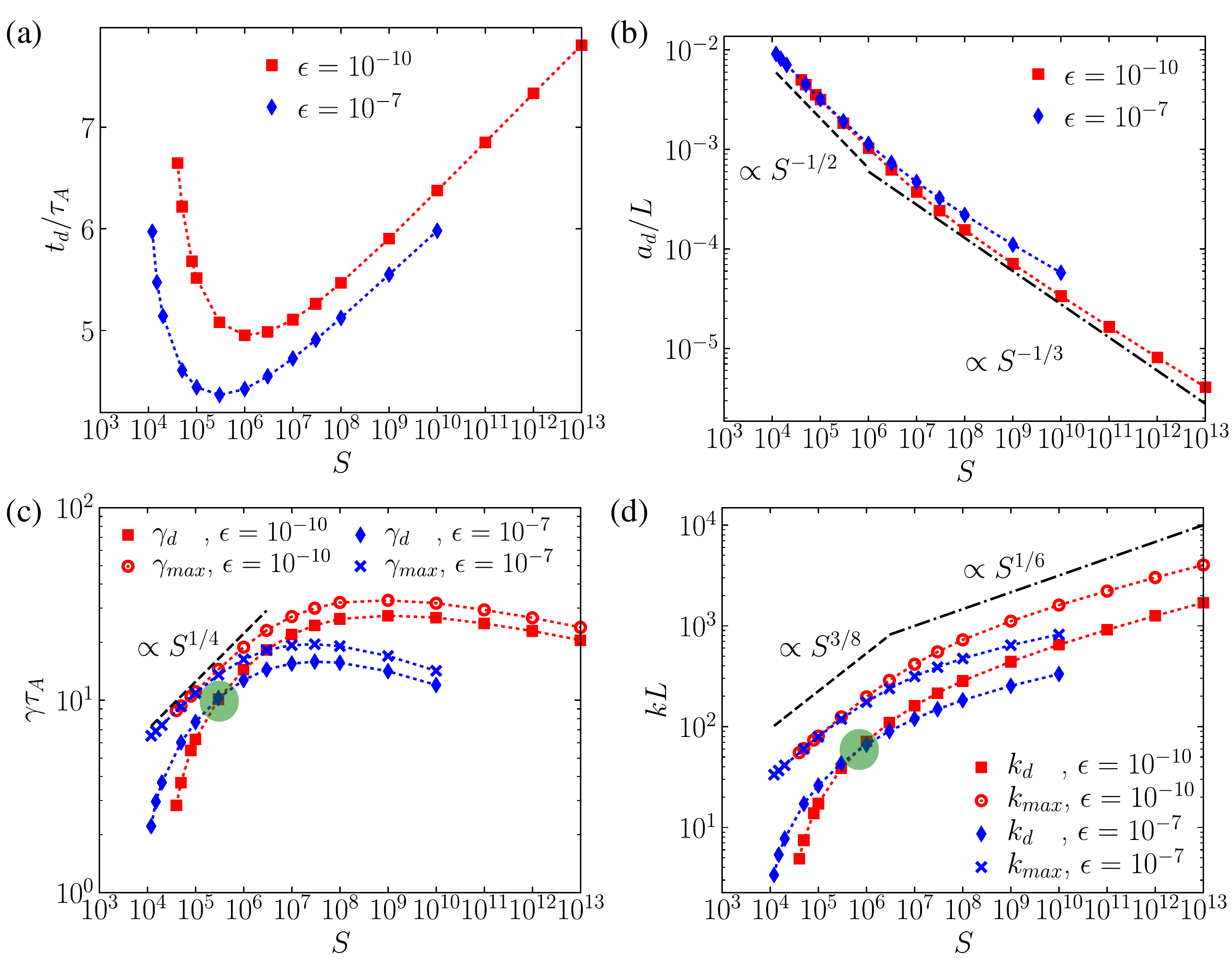}
\par\end{centering}
\caption{Scalings of $t_{d}$, $a_{d}$, $k_{d}$, and $\gamma_{d}$ with respect
to $S$ from the model with $f_{0}(k)=\epsilon$, for $\epsilon=10^{-7}$
and $10^{-10}$. \textcolor{black}{Green circles in panels (c) and
(d) highlight the crossing between two $\text{\ensuremath{\gamma}}_{d}(S)$
curves and two $k_{d}(S)$ curves with different $\epsilon$. }\label{fig:Scalings-model-1}}
\end{figure*}
\begin{figure*}
\begin{centering}
\includegraphics[scale=0.63]{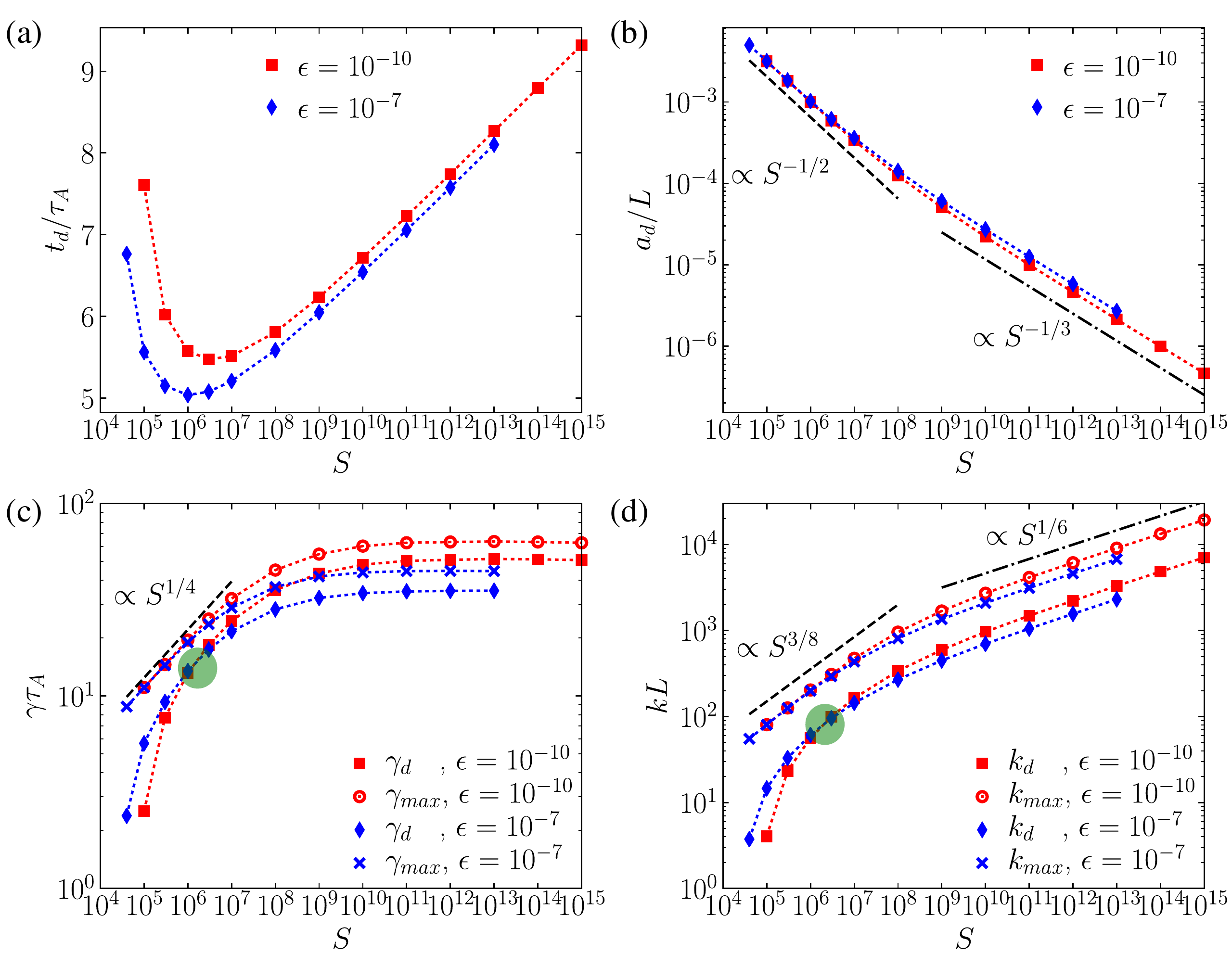}
\par\end{centering}
\caption{Scalings of $t_{d}$, $a_{d}$, $k_{d}$, and $\gamma_{d}$ with respect
to $S$ from the model, with an initial noise $f_{0}=\epsilon(k/k_{min})^{-1}$,
for $\epsilon=10^{-7}$ and $10^{-10}$. \textcolor{black}{Green circles
in panels (c) and (d) highlight the crossing between two $\text{\ensuremath{\gamma}}_{d}(S)$
curves and two $k_{d}(S)$ curves with different $\epsilon$.}\label{fig:Scalings-model-3}}
\end{figure*}
\begin{figure}
\begin{centering}
\includegraphics[scale=0.65]{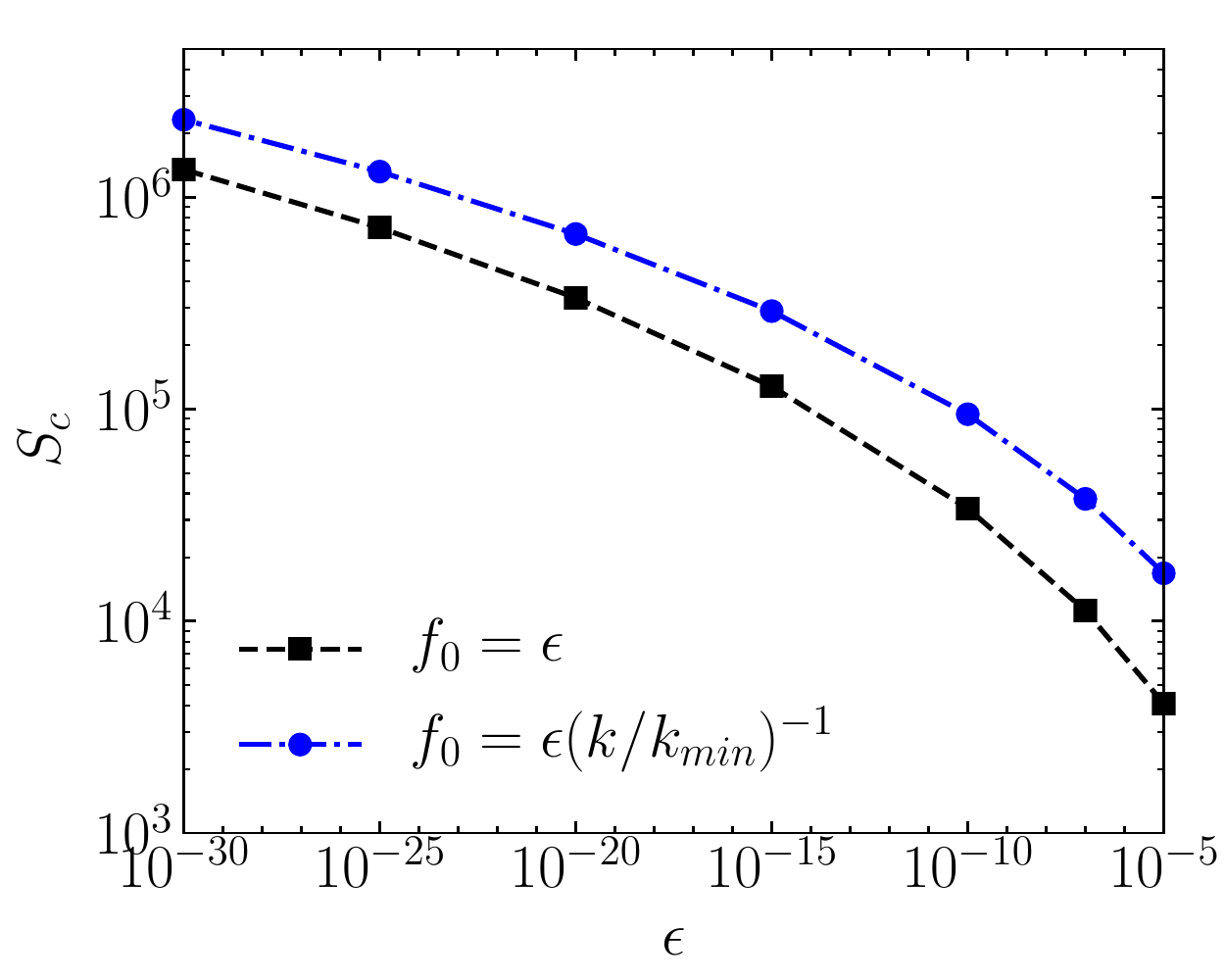}
\par\end{centering}
\caption{Critical Lundquist number $S_{c}$ as a function of initial noise,
for $f_{0}(k)=\epsilon$ and $f_{0}(k)=\epsilon(k/k_{min})^{-1}$.
 \label{fig:Critical-Lundquist-number}}
\end{figure}

To apply the model equation, an initial condition must be prescribed
for the fluctuation amplitude $f(k)$. For the time evolution of a
current sheet given in the precious paragraph, we can solve Eq.~(\ref{eq:linear-growth})
using the method of characteristics. Along a characteristic following
the stretching of wavenumber
\begin{equation}
k=k_{0}e^{-t/\tau_{A}},\label{eq:characteristic}
\end{equation}
the solution is
\begin{equation}
f(k,t)=f_{0}(k_{0})\exp\left(\int_{0}^{t}\gamma(k(t'),a(t'))dt'-\frac{t}{2\tau_{A}}\right),\label{eq:solution}
\end{equation}
where $f_{0}(k)$ is the initial condition. We numerically integrate
Eq.~(\ref{eq:solution}) to obtain the solution. To fix the idea,
we set the initial half-width $a_{0}=L/\pi$ such that the longest
mode allowed in the model with $k=k_{min}$ is marginally unstable.
We identify the dominant wavenumber by the peak of $f(k)$ and determine
the disruption time by the criterion that the island size equals the
inner layer width, using the same procedure outlined in Sec.~\ref{sec_Diagnostics}
to calculate the island width. 

We first assume the initial condition to be a constant value, $f_{0}(k)=\epsilon$.\footnote{Strictly speaking, a constant $f_{0}(k)=\epsilon$ is not permitted,
because $\int_{k_{min}}^{\infty}f_{0}^{2}dk$ diverges. This problem
can be overcome if we assume that $f_{0}(k)=\epsilon$ for a broad
range of $k$ but approaches zero for very high-$k$ modes. As long
as those high-$k$ modes never become unstable to the tearing instability
(following mode-stretching), the results we obtain remain unchanged. } As an example, the solution for the case $S=10^{8}$ and $\epsilon=10^{-10}$
is shown in Figure \ref{fig:model-solution}. Here panel (a) shows
the linear growth rate as a function of time and $k$, and panel (b)
shows the solution obtained from integrating Eq.~(\ref{eq:solution}).
The time integration concludes when the disruption condition is met.
Also shown in these panels are the wavenumber of the fastest growing
mode and that of the dominant mode at each instant of time. The wavenumber
of the dominant mode is found to be smaller than that of the fastest
growing mode, which is consistent with the simulation results. The
reason that the dominant mode does not coincide with the fastest growing
mode at disruption can be seen from their corresponding characteristics,
shown as white dotted lines in panel (a). Because the exponent in
Eq.~(\ref{eq:solution}) is given by the integration of the linear
growth rate over the characteristic, the amplification of a mode is
determined by the entire history. As can be seen from Fig.~\ref{fig:model-solution}(a),
although the fastest growing mode has a higher growth rate at the
final moment before disruption, the dominant mode actually becomes
unstable earlier and grows faster during most of the time. 

The scalings of $t_{d}$, $a_{d}$, $\gamma_{d}$, and $k_{d}$ from
the model are shown in Figure \ref{fig:Scalings-model-1} for $\epsilon=10^{-7}$
and $10^{-10}$. These scalings can be compared with results from
simulations shown in Figure \ref{fig:Scaling-S}. Qualitative agreement
between the two is evident, including the non-monotonicity of $t_{d}$
with respect to $S$, the transition in scaling behaviors as $S$
increases, the deviation between the fastest growing mode and the
dominant mode, and the effects of noise amplitude on various quantities.
Notably, the model is able reproduce the crossing of the two $\gamma_{d}(S)$
curves (highlighted with a green circle in panel (c)), namely, while
$\gamma_{d}$ is higher for the lower initial noise case at high $S$,
it is the opposite at low $S$. The reason for the crossing can be
attributed to the mode-stretching effect due to the outflow jets.
The stretching effect is more prominent at lower $S$, as reflected
in the larger difference between $\gamma_{max}$ and $\gamma_{d}$
there. Moreover, when $S$ is sufficiently low, the amplitudes of
unstable modes cannot grow sufficiently to disrupt the current sheet
before being advected out. Therefore, there is a critical Lundquist
number $S_{c}$ below which disruption does not occur. The critical
Lundquist number $S_{c}$ depends on the initial noise amplitude;
the lower the noise amplitude is, the higher $S_{c}$ becomes.  Because
$\gamma_{d}$ becomes very low as $S$ approaches $S_{c}$ (precisely
$\gamma_{d}\to\gamma(\pi/L,a_{SP})$), the crossing of $\gamma_{d}(S)$
curves with different $\epsilon$ at low $S$ is simply a consequence
of $S_{c}$ being a decreasing function with respect to $\epsilon$.
A similar crossing is also observed between the dominant wavenumber
$k_{d}(S)$ curves, shown in panel (d), because $k_{d}\to\pi/L$ as
$S$ approaches $S_{c}$.

\begin{figure*}
\begin{centering}
\includegraphics[scale=0.63]{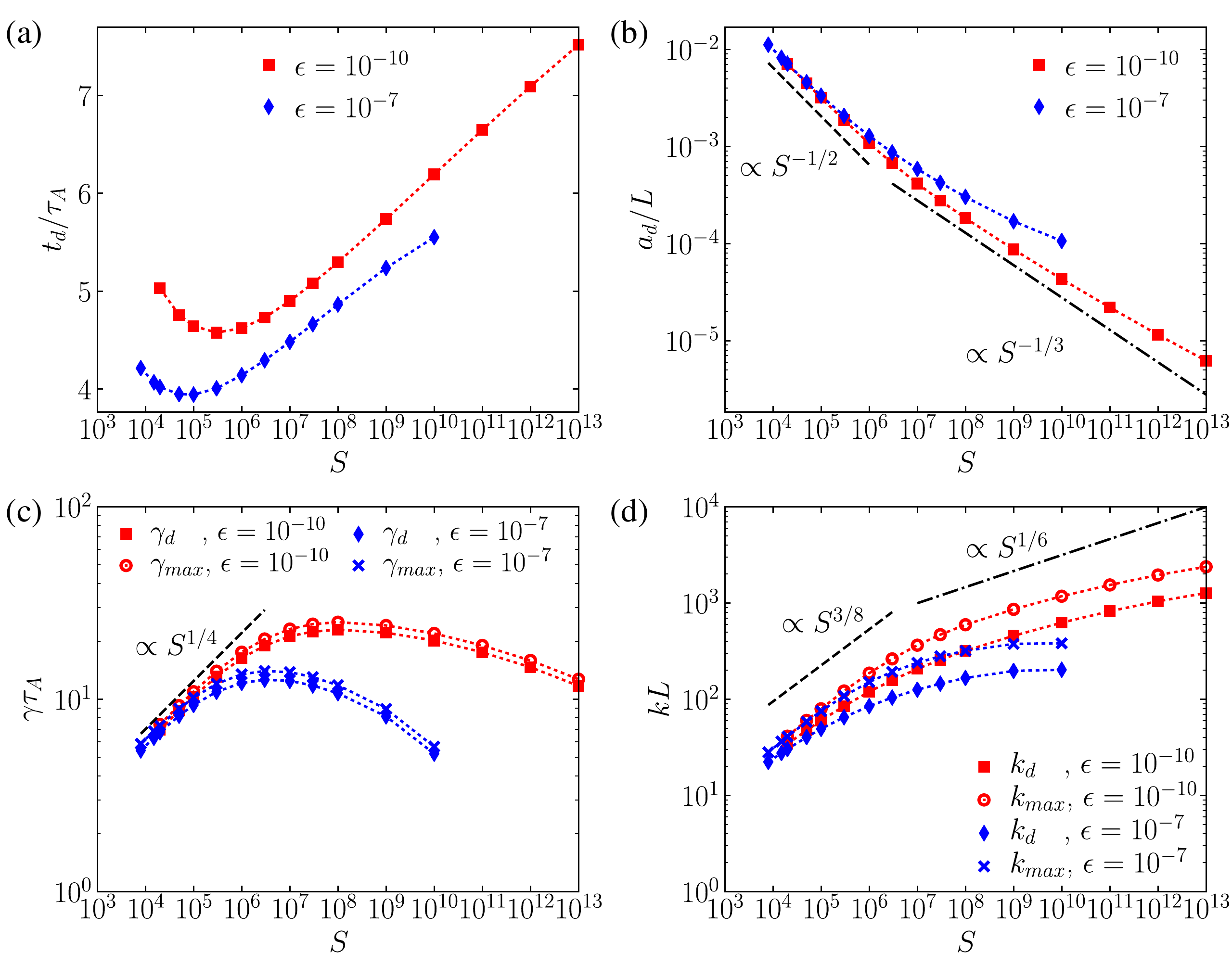}
\par\end{centering}
\caption{Scalings of $t_{d}$, $a_{d}$, $k_{d}$, and $\gamma_{d}$ with respect
to $S$ from the model, with the effect of outflow jets turned off.\label{fig:Scalings-model-2}}
\end{figure*}
Another notable outcome from the model is the non-monotonicity of
$\gamma_{d}$ and $\gamma_{max}$ with respect to $S$. At low $S$,
the current sheet half-width $a_{d}\simeq a_{SP}$ at disruption and
$\gamma_{max}\sim S^{1/4}$ is an increasing function with respect
to $S$. At high $S$, $\gamma_{max}$ becomes a weakly deceasing
function with respect to $S$, and $\gamma_{d}$ follows the same
trend. Recall that the condition $a_{d}/L\propto S^{-1/3}$ gives
a constant value of $\gamma_{max}$ independent of $S$, and $k_{max}L\propto S^{1/6}$.
Because $\gamma_{max}$ is not exactly constant but is weakly decreasing
with increasing $S$, the scalings of $a_{d}$ and $k_{max}$ are
close to, but not exactly $a_{d}/L\propto S^{-1/3}$ and $k_{max}L\propto S^{1/6}$
($k_{d}$ approximately follows the same trend as $k_{max}$ at high
$S$).  This weakly decreasing scaling for $\gamma_{max}$ (or $\gamma_{d}$)
at high $S$ has not been observed in numerical simulation yet. The
reason could be that the numerical simulations have not been carried
out at sufficiently high values of $S$ to reveal this behavior.

We have tested the model with different initial conditions. As an
example, Figure \ref{fig:Scalings-model-3} shows the scalings with
$f_{0}=\epsilon(k/k_{min})^{-1}$. Comparing with Fig.~\ref{fig:Scalings-model-1},
the overall trends are rather similar between the two, even though
the actual quantitative values can be quite different. This qualitative
similarity in scaling relations remains the case for other initial
conditions we have tried. Likewise, we have also tested the effect
of noise source by setting a floor to the fluctuation amplitude, assumed
to be the same as the initial noise $f_{0}(k)$, and similar scaling
relations are obtained. 

As we have discussed, the crossings between $\gamma_{d}(S)$ and $k_{d}(S)$
curves with respect to different values of initial noise amplitude
$\epsilon$ is a consequence of $S_{c}$ being a decreasing function
with respect to $\epsilon$. The dependence of $S_{c}$ on $\epsilon$
is fairly weak. For the two types of initial condition we have tried,
the critical value $S_{c}$ as a function of $\epsilon$ is shown
in Figure \ref{fig:Critical-Lundquist-number}. In both cases, the
initial noise amplitude $\epsilon$ is varied over 25 orders of magnitude
but the corresponding $S_{c}$ changes no more than 3 orders of magnitude. 

We can assess the effect of outflow jets by switching off the $v_{x}'$
terms in Eq.~(\ref{eq:linear-growth}). The resulting scalings are
shown in Fig.~\ref{fig:Scalings-model-2}. Overall, the scalings
are similar to the scalings with outflow in Fig.~\ref{fig:Scalings-model-1}.
The most significant differences between the two appear in the low-$S$
regime. In the case without outflow, the difference between $\gamma_{d}$
and $\gamma_{max}$, as well as that between $k_{d}$ and $k_{max}$,
become smaller in the low-$S$ regime, as opposed to the case with
outflow, where the differences become larger at low $S$. We also
do not find the crossing between curves with different $\epsilon$
when outflow is not included in the model. In fact, without the effect
of outflow, the critical Lundquist number will be determined by the
condition that the longest wavelength mode allowed by the current
sheet length $L$ is marginally unstable when the current sheet width
is the Sweet-Parker width. This condition gives a critical value $S_{c}=\pi^{2}$,
which is independent of the noise amplitude. This value of $S_{c}$
substantially underestimates the actual value of $S_{c}$.

\section{From Disruption to Nonlinear Saturation and Onset of Fast Reconnection }

\label{sec:nonlinear}
\begin{figure*}
\begin{centering}
\includegraphics[scale=0.63]{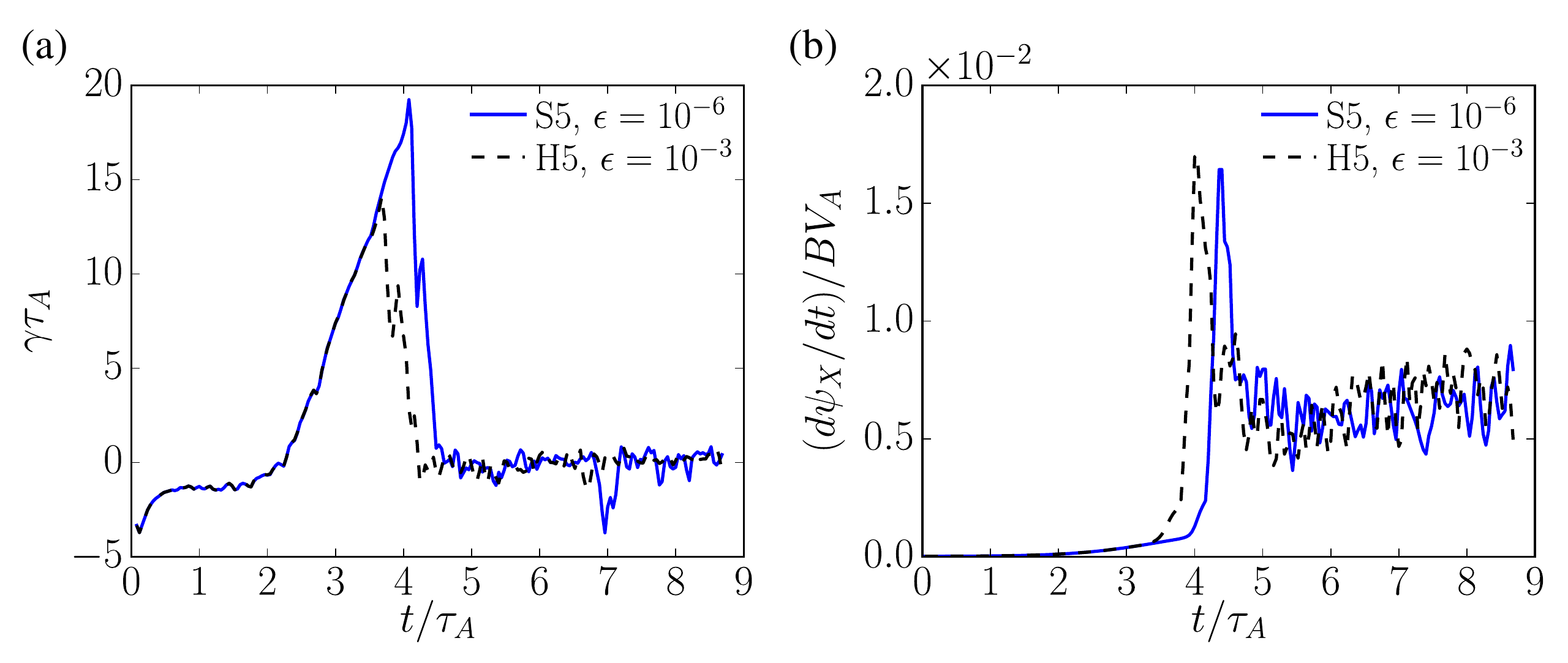}
\par\end{centering}
\caption{Comparison of time histories of (a) linear growth rate and (b) reconnection
rate, for Run S5 (with an initial noise level $\epsilon=10^{-6}$)
and Run H5 ($\epsilon=10^{-3}$). The initial noise level affects
the disruption time and the linear growth rate at disruption, but
the reconnection rate after nonlinear saturation remains approximately
the same. \label{fig:rate-comparison} }
\end{figure*}

Figure \ref{fig:Run-S5}(a) shows that the reconnection rate rises
rapidly after the disruption time $t_{d}$, indicating that current
sheet disruption triggers the transition from slow to fast reconnection.
Now we have established that the linear growth rate and dominant wavenumber
at disruption have nontrivial dependences on various conditions, including
the thinning process, the Lundquist number $S$, the initial noise
level and even the spectrum of noise. An important question is, do
these conditions affect the reconnection process after the plasmoid
instability reaches nonlinear saturation? A full assessment of this
question is beyond the present scope and left to future work. To investigate
the effect of initial noise level on reconnection rate, we run the
two cases S5 and H5 for extended periods well into the nonlinearly
saturated phase. The resulting time-histories of growth rate and reconnection
rate are shown in Fig.~\ref{fig:rate-comparison}. Panel (a) shows
that current sheet disruption takes place earlier with a lower growth
rate in Run H5, where the initial noise level is three orders of magnitude
higher than that of Run S5. Because current sheet disruption marks
the beginning of onset of fast reconnection, the onset also occurs
earlier in Run H5. However, as can be seen from panel (b), the reconnection
rates after nonlinear saturation are essentially the same for the
two runs. Therefore, we can conclude that reconnection rate after
saturation is nearly unaffected by the noise level and the condition
at the onset. This conclusion is also in line with the earlier study
by \citet{HuangB2010}, where the dependence of averaged reconnection
rate on external forcing amplitude is found to be fairly weak. 

\begin{figure*}
\begin{centering}
\includegraphics[scale=0.65]{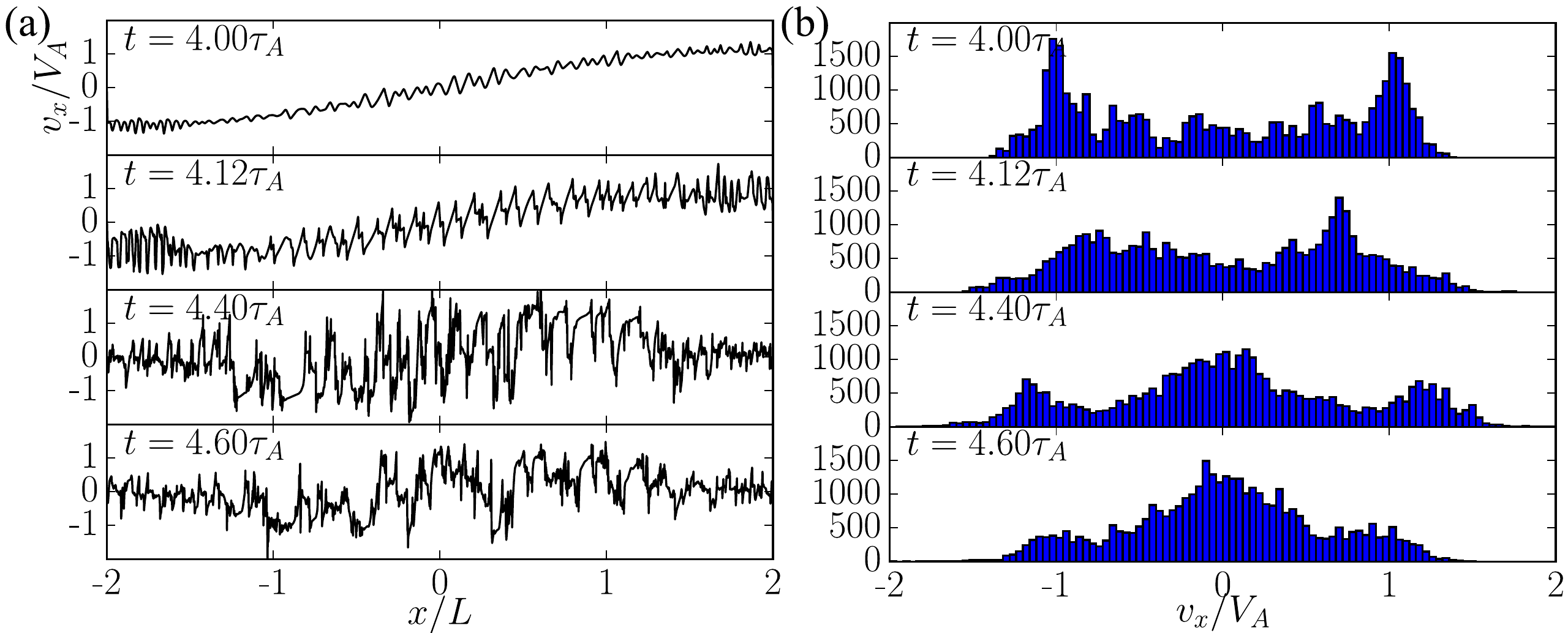}
\par\end{centering}
\caption{(a) Time sequence of the outflow ($v_{x}$) profile along $z=0$ from
disruption ($t=4\tau_{A}$) to nonlinear saturation ($t=4.6\tau_{A}$)
for Run S5. (b) The corresponding histogram of the outflow velocity.
\label{fig:velocity-profile}}
\end{figure*}

From an observational point of view, the period between current sheet
disruption and nonlinear saturation is especially interesting. During
this period, the outflow profile changes from a nearly monotonic,
Sweet-Parker-like profile, to a highly fluctuating profile, as shown
in Figure \ref{fig:velocity-profile}(a). Accordingly, the histogram
of the outflow profile changes from double-peak-shaped to triangle-shaped,
as shown in Figure \ref{fig:velocity-profile}(b). A triangle-shaped
histogram is a natural consequence of the outflow profile being highly
fluctuating. To understand that, consider an outflow profile of the
form $v_{x}=V_{A}\sin(\pi x)+v_{f}\sin(kx)$, where $-1\le x\le1$
and $k\gg1$; the first term $V_{A}\sin(\pi x)$ mimics the Sweet-Parker-like
outflow profile, and the second term $v_{f}\sin(kx)$ mimics short-wavelength
fluctuations. If the fluctuation is much smaller than the background,
\emph{i.e.} $v_{f}\ll V_{A}$, the histogram is double-peak-shaped.
On the contrary, if the fluctuation is of the order of the background,
\emph{i.e.} $v_{f}\simeq V_{A}$, which is the case in the saturated
phase, then the histogram is triangle-shaped. The transition from
a double-peak-shaped histogram to triangle-shaped histogram can lead
to observable effects through spectroscopy of emission spectral line
profiles, because the observed spectral line profile is a convolution
of the Doppler-shifted line profile along the line-of-sight direction.
For reconnection events with outflows along the line-of-sight direction,
the histogram of the outflow profile can be viewed as a proxy for
the shape of the line profile. A triangle-shaped line profile has
been reported in IRIS (Interface Region Imaging Spectrograph \citep{DePontieuTL2014})
observations of the Si IV line during transition region explosive
events \citep{InnesGHB2015}. With the high-cadence mode of IRIS spectrograph,
it is possible to observe transitions from a double-peak-shaped line
profile to a triangle-shaped line profile in transition region explosive
events \citep{Guo2017}. If such transitions can be shown to associate
with brightening of emission spectral line, that will provide observational
evidence for transition from slow to fast reconnection due to current
sheet disruption.

\section{Discussion and Conclusion}

\label{sec:conclusion}In this work, we have carried out a detailed
study of the development of plasmoid instability in an evolving current
sheet and the onset of fast reconnection. We confirm that the Sweet-Parker
width can be approached only when the Lundquist number is relatively
low. At high values of the Lundquist number, more characteristic of
the solar and astrophysical environments, the current sheet breaks
apart at a width much larger than the Sweet-Parker width \citep{PucciV2014,UzdenskyL2016,ComissoLHB2016}.
However, the level of noise in the system also plays an important
role. For a system with lower noise amplitude, the Sweet-Parker width
can be approached at a higher Lundquist number and consequently the
linear growth rate becomes higher. As such, the linear growth rate
and the dominant wavenumber at current sheet disruption exhibit complex
dependences on numerous conditions, such as the current sheet evolving
process, the Lundquist number, noise level, and even the spectrum
of the noise. \emph{Under the condition that the Lundquist number
$S$ is the only varying parameter while everything else remains fixed},
we do find that the growth rate at disruption becomes slowly varying
with respect to $S$, as long as $S$ is sufficiently high such that
the Sweet-Parker scaling $a_{d}/L\sim S^{-1/2}$ is no longer valid.
Because $a_{d}/L\propto S^{-1/3}$ is the condition for $\gamma_{max}\tau_{A}$
to be independent of $S$, in the high-$S$ regime, the scalings $a_{d}/L\propto S^{-1/3}$
and $k_{d}L\propto S^{1/6}$ are approximately realized. While these
scaling relations appear to be similar to the suggestion by \citet{PucciV2014},
Pucci \& Velli actually imposed a more stringent condition $a_{d}/L=S^{-1/3}$,
leading to the condition that $\gamma_{max}\tau_{A}\simeq0.623$ at
disruption. Contrary to their suggestion that $\gamma_{max}\tau_{A}=O(1)$
at disruption, we generally find the linear growth rate at disruption
to be substantially higher than $1/\tau_{A}$. In fact, the time when
$\gamma_{max}\tau_{A}=O(1)$ is approximately when the overall amplitude
of fluctuations starts to grow. This hardly comes as a surprise, since
the linear growth must be substantial on time scales pertaining to
the evolving background for linear theory to be even applicable. 
In the present case the relevant time scales are the advection and
the current sheet evolution time scales, both of the order of $\tau_{A}$.
That means $\gamma_{max}\tau_{A}>O(1)$ is required. Because $\gamma_{max}\propto a^{-3/2}$,
using the criterion $\gamma_{max}\tau_{A}\simeq1$ for disruption
will substantially overestimate the disruption half-width $a_{d}$,
even if the scaling relation $a_{d}/L\propto S^{-1/3}$ appears to
approximately hold; likewise, using the criterion $\gamma_{max}\tau_{A}\simeq1$
will substantially underestimate the fastest growing wavenumber $k_{max}$.
For example, using $\gamma_{max}\tau_{A}\simeq1$ instead of $\gamma_{max}\tau_{A}\simeq20$
as we typically observe in simulations will overestimate $a_{d}$
by a factor of approximately $7$ and underestimate the fastest growing
wavenumber $k_{max}$ by a factor of approximately 12. Incidentally,
if $k_{max}$ is used to estimate the dominant wavenumber $k_{d}$
at disruption, as is often assumed, then the error of underestimating
$k_{max}$ is partially offset by the fact that the dominant wavenumber
$k_{d}$ is actually smaller than $k_{max}$.

Our simulation study differs from the earlier study by \citet{TeneraniVRP2015}
in an important aspect. In our study the current sheet thinning is
self-consistently determined by MHD equations, where Alfv\'enic jets
develops naturally during the process. This thinning process approaches
the Sweet-Parker width as the asymptotic state if the plasmoid instability
does not set in. On the other hand, \citet{TeneraniVRP2015} employ
an extra term in the induction equation to drive the thinning, therefore
the asymptotic width is set by the driving term, and there is no significant
outflow before the instability sets in. In three out of four simulations
(run 1 \textendash{} run 3) in their study, the asymptotic width is
set to be $LS^{-1/3}$, exactly as their theory predicts; in the fourth
case (run 4), the asymptotic width is set to be the Sweet-Parker width
$LS^{-1/2}$ {[}see Table 1 in \citep{TeneraniVRP2015}{]}. It is
particularly revealing to compare their run 1 and run 4, both have
exactly the same parameters except the asymptotic width. While the
current sheet in run 1 disrupts at $a(\tau_{nl})/L\simeq S^{-1/3}$
($\tau_{nl}$ in their Table 1 is equivalent to $t_{d}$ in our paper
and $a(\tau_{nl})$ is equivalent to $a_{d}$ in our paper), when
the current sheet is allowed to go thinner in run 4, the disruption
occurs at an earlier time but the current sheet width is approximately
4 times thinner. It is worth noting that \citet{TeneraniVRP2015}
use a fixed value $S=10^{6}$ in all their simulations, thereby precluding
the possibility of obtaining scaling relations with respect to $S$.

The phenomenological model presented in Sec.~\ref{sec:Phenomenological-Model}
qualitatively reproduces many features obtained from direct numerical
simulations, including the non-monotonicity of $t_{d}$ and changes
of scaling behaviors as $S$ increases. It also clarifies why the
dominant wavenumber $k_{d}$ differs from the fastest growing wavenumber
$k_{max}$. An important feature of this model is its capability to
predict a critical Lundquist number $S_{c}$ for a given current sheet
evolving process and an initial noise spectrum. The critical Lundquist
number $S_{c}$ is found to have a weak dependence on the noise amplitude.
As shown in Fig.~\ref{fig:Critical-Lundquist-number}, for the two
types of initial perturbation, the critical value $S_{c}$ falls in
the range from $S_{c}\simeq10^{3}$ to $10^{6}$ for a change in the
noise amplitude $\epsilon$ over 25 orders of magnitude. This result
makes the qualitative argument in \citep{HuangB2013} more precise,
and provides new insight to some empirical knowledge as well as controversy
among researchers. The often quoted critical Lundquist number $S_{c}$
is approximately $10^{4}$, but this is not a clear-cut value. For
example, a critical value $S_{c}$ as low as $10^{3}$ \citep{ShenLMR2013}
as well as clean simulations remaining stable up to $S=2\times10^{5}$
\citep{NgR2011} have been reported. These seemingly conflicting results
become comprehensible with the realization that the critical value
$S_{c}$ depends on noise level. Noise can enter a simulation in various
ways, such as explicit seeding, discretization errors of numerical
schemes, and the implementation of boundary conditions. The simulation
setup of \citet{NgR2011} using a pseudospectral code in a doubly-periodic
box without seeded noise is among the cleanest of all scenarios, which
accounts for its stability. However, since natural systems always
have some noise and the Lundquist numbers are typically much higher,
it is unlikely that such systems can remain stable. As we have shown
in Sec.~\ref{sec:nonlinear}, the level of noise affects the disruption
time, but has a negligible effect on reconnection rate once the instability
reaches nonlinear saturation (Fig.~\ref{fig:rate-comparison}).

Our results can also be compared with two recent theoretical approaches
by \citet{UzdenskyL2016} and \citet{ComissoLHB2016}. The analysis
of \citet{UzdenskyL2016} assumes that the current sheet width essentially
freezes once the maximum linear growth rate of all modes that fit
into the current sheet (\emph{i.e.} $k>\pi/L$) reaches $1/\tau_{dr}$,
where $\tau_{dr}$ is the current sheet evolution time scale. Their
analysis concludes that the fastest growing mode at that time will
be the dominant mode that enters the nonlinear regime and eventually
disrupts the current sheet. As such, in the end the criterion of \citet{UzdenskyL2016}
($\gamma_{max}\tau_{dr}=1$) is similar to that of \citet{PucciV2014}
($\gamma_{max}\tau_{A}=O(1)$), except that the Alfv\'enic time scale
$\tau_{A}$ in the latter is replaced by $\tau_{dr}$. In the context
of the present study, the two criteria are practically the same since
$\tau_{dr}\simeq\tau_{A}$. The time when $\gamma_{max}\tau_{A}=1$
is slightly earlier than $t=t_{g}$ in our simulations. As can be
seen from Table \ref{tab:run_table}, the current sheet width evolves
quite significantly from $t=t_{g}$ to $t=t_{d}$ (the actual difference
depends on initial noise level, as does $t_{d}$) and consequently
the fastest growing wavenumber also changes substantially (see Fig.~\ref{fig:Run-S5});
therefore, the basic assumption of \citet{UzdenskyL2016} is not borne
out by our simulations.  It is worth noting that \citet{UzdenskyL2016}
use island width $w$ exceeding the current layer width $a$ as the
definition of disruption, which is different from our definition of
$w$ exceeding the inner layer width $\delta$; this difference does
not affect the discussion here. On the other hand, the analysis of
\citet{ComissoLHB2016} starts from a principle of least time, which
assumes that the dominant mode at disruption is the one that takes
the least time to reach the condition $w=\delta$, for a given initial
spectrum $w_{0}(k)$. The scalings obtained from the principle of
least time is very similar to the scalings obtained with our model.
The only significant difference is that the effect of outflow is not
taken into account in \citet{ComissoLHB2016}. The reader can compare
Fig.~\ref{fig:Scalings-model-2}(c)(d) (where the effect of outflow
is turned off) in this paper and Fig.~1 and Fig.~2 in \citet{ComissoLHB2016}
to see the similarities. As we have discussed, the effect of outflow
is most pronounced when the Lundquist number is relatively low; therefore,
the theory of \citet{ComissoLHB2016} and the model here practically
give very similar scalings in the asymptotic limit of large $S$.
For an exponentially thinning current sheet, \citet{ComissoLHB2016}
give precisely the logarithmic multiplying factors to the leading
order power-law scalings $a_{d}/L\propto S^{-1/3}$ and $k_{d}L\propto S^{1/6}$
{[}i.e. the scalings are of the form $\sim S^{\alpha}(\log f(S,w_{0}))^{\beta}$,
see their Eqs.~(17) and (18){]}, which account for the deviations
from power-laws in Fig.~\ref{fig:Scalings-model-1}(b)(d). It is
important to note that these leading order power-law scalings are
not universal, as the power indices will be different if the thinning
is algebraic rather than exponential {[}see their Eq.~(24){]}.

We conclude this paper by applying our findings on current sheet disruption
to reconnection in the solar atmosphere. As a first application, consider
the example of post-CME current sheet in \citep{GuoBH2013}. The current
sheet length is estimated as $L\simeq3\times10^{11}\,\text{cm}$,
and the Alfv\'en speed can be estimated from the speed of moving
blobs as $V_{A}\simeq2\times10^{7}\,\text{cm/s}$. Assuming a density
of $10^{10}\,\text{cm}^{-3}$ and a temperature of $10^{6}\,\text{K}$
for typical solar coronal condition, the Lundquist number $S\simeq3\times10^{14}$.
The actual disruption condition will depend on the initial noise level,
which is not easy to know. If we assume a typical value $\gamma_{max}\tau_{A}\simeq20$
at disruption, that gives an inverse aspect ratio $a_{d}/L\simeq1.5\times10^{-6}$,
\emph{i.e.} $a_{d}\simeq4.5\times10^{5}\,\text{cm}$. Assuming a typical
dominant wavenumber $k_{d}\simeq k_{max}/4$, we estimate $k_{d}L\simeq1500$,
and the corresponding inner layer width $\delta\simeq5000\,\text{cm}$.
Since the inner layer width is much larger than the ion inertial length
$d_{i}\simeq200\,\text{cm}$ and ion gyroradius $\rho_{i}\simeq60\,\text{cm}$,
the resistive MHD model remains valid until disruption. However, nonlinearly
the fractal-like cascade will break the current sheet to finer scales.
Within the framework of resistive MHD, the cascade will stop when
the secondary current sheets become marginally stable \citep{HuangB2010}.
This condition gives an estimate for the secondary current sheet width
$\delta_{c}\simeq a_{SP}\sqrt{S_{c}/S}$. Taking a typical value $S_{c}=10^{4}$
yields $\delta_{c}\simeq0.1\,\text{cm}$, which is much smaller than
either $d_{i}$ or $\rho_{i}$, therefore collisionless effects must
be taken into account at nonlinear saturation \citep{DaughtonRAKYB2009,ShepherdC2010,HuangBS2011}.
As another application, consider the transition region explosive events
reported in \citep{InnesGHB2015}. If we estimate the length scale
$L\simeq10^{8}\text{\,cm}$ and the Alfv\'en speed $V_{A}\sim2\times10^{7}\text{\,cm/s}$,
then $\tau_{A}\simeq5\text{\,s}$. For a temperature $10^{5}\,\text{K}$
and a density $10^{12}\,\text{cm}^{-3}$, the Lundquist number in
transition region can be estimated as $S\simeq5\times10^{9}$ based
on Spitzer resistivity. Again, assuming $\gamma_{max}\tau_{A}\simeq20$
and $k_{d}\simeq k_{max}/4$ at disruption, we obtain $a_{d}\simeq5800\,\text{cm}$,
$k_{d}L\simeq250$, and $\delta\simeq400\,\text{cm}$. Since $\delta$
is much greater than $d_{i}\simeq20\,\text{cm}$ and $\rho_{i}\simeq3\,\text{cm}$,
resistive MHD is applicable at disruption. The secondary current sheet
at nonlinear saturation based on resistive MHD is estimated to be
$\delta_{c}\simeq2\,\text{cm}$, which is smaller than $d_{i}$ by
an order of magnitude and is comparable to $\rho_{i}$. Therefore
collisionless effects will not be negligible at nonlinear saturation
for anti-parallel reconnection, but may only be marginally important
for component reconnection. How collisionless effects at small scales
affect nonlinear saturation in systems with very large $L/d_{i}$
or $L/\rho_{i}$ remains an outstanding open question. 

Although the present study is restricted to a resistive MHD description,
the overall framework can be readily generalized to incorporate other
effects. The phenomenological model can be adapted, provided that
the linear dispersion relation of tearing instability is available,
\emph{e.g.} in visco-resistive \citep{ComissoG2016,ComissoLHB2017}
and resistive Hall \citep{BaalrudBHG2011} regimes. These effects
will be considered in future studies. Furthermore, the current sheet
evolving process investigated in this study, where the upstream magnetic
field remains approximately constant and the current sheet width decreases
exponentially in time, is by no means universal, as the current sheet
evolution is largely determined by the global configuration of the
system. An interesting example is the scenario of reconnection in
a layer embedded within a broader outer current sheet, considered
by \citet{CassakD2009}. In this scenario, the upstream magnetic field
of the embedded layer increases in time; the thinning of the layer,
while remains to be studied in detail, may also not be exponential
in time. This setup could be employed to investigate how different
evolution processes may affect the scalings and further test our phenomenological
model. More general situations, where the current sheet length evolves
in time, should also be investigated in the future.

\acknowledgements{Beneficial discussion with Dr.~Lijia Guo on IRIS observation is
gratefully acknowledged. We thank the anonymous referee for many constructive
suggestions. We also thank Dr.~Nick Murphy for useful conversations
when the paper was under revision. This work is supported by the National
Science Foundation, Grant Nos.~AGS-1338944 and AGS-1460169, and the
Department of Energy, Grant No.~DE-SC0016470. Simulations were performed
with supercomputers at the Oak Ridge Leadership Computing Facility
and the National Energy Research Scientific Computing Center.}


\begin{thebibliography}{}
\expandafter\ifx\csname natexlab\endcsname\relax\def\natexlab#1{#1}\fi
\providecommand{\url}[1]{\href{#1}{#1}}

\bibitem[{{Alfv{\'e}n}(1943)}]{Alfven1943}
{Alfv{\'e}n}, H. 1943, Arkiv f\"or matematik, astronomi och fysik, 29, 1

\bibitem[{Baalrud {et~al.}(2011)Baalrud, Bhattacharjee, Huang, \&
  Germaschewski}]{BaalrudBHG2011}
Baalrud, S.~D., Bhattacharjee, A., Huang, Y.-M., \& Germaschewski, K. 2011,
  Phys. Plasmas, 18, 092108

\bibitem[{Bhattacharjee(2004)}]{Bhattacharjee2004}
Bhattacharjee, A. 2004, Annu. Rev. Astron. Astrophys., 42, 365

\bibitem[{Bhattacharjee {et~al.}(2009)Bhattacharjee, Huang, Yang, \&
  Rogers}]{BhattacharjeeHYR2009}
Bhattacharjee, A., Huang, Y.-M., Yang, H., \& Rogers, B. 2009, Phys. Plasmas,
  16, 112102

\bibitem[{Biskamp(1993)}]{Biskamp1993}
Biskamp, D. 1993, Nonlinear Magnetohydrodynamics (Cambridge University Press)

\bibitem[{Biskamp(2000)}]{Biskamp2000}
---. 2000, Magnetic Reconnection in Plasmas (Cambridge University Press)

\bibitem[{Boyd(2001)}]{Boyd2001}
Boyd, J.~P. 2001, Chebyshev and Fourier Spectral Methods, 2nd edn. (Dover
  Publications, Inc.)

\bibitem[{Cassak \& Drake(2009)}]{CassakD2009}
Cassak, P.~A., \& Drake, J.~F. 2009, Astrophys. J. Lett., 707, L158

\bibitem[{Cassak {et~al.}(2005)Cassak, Shay, \& Drake}]{CassakSD2005}
Cassak, P.~A., Shay, M.~A., \& Drake, J.~F. 2005, Phys. Rev. Lett., 95, 235002

\bibitem[{Cassak {et~al.}(2009)Cassak, Shay, \& Drake}]{CassakSD2009}
---. 2009, Phys. Plasmas, 16, 120702

\bibitem[{Comisso \& Grasso(2016)}]{ComissoG2016}
Comisso, L., \& Grasso, D. 2016, Phys. Plasmas, 032111

\bibitem[{Comisso {et~al.}(2015)Comisso, Grasso, \& Waelbroeck}]{ComissoGF2015}
Comisso, L., Grasso, D., \& Waelbroeck, F.~L. 2015, Phys. Plasmas,
  doi:10.1063/1.4918331

\bibitem[{Comisso {et~al.}(2016)Comisso, Lingam, Huang, \&
  Bhattacharjee}]{ComissoLHB2016}
Comisso, L., Lingam, M., Huang, Y.-M., \& Bhattacharjee, A. 2016, Phys.
  Plasmas, 23, 100702

\bibitem[{Comisso {et~al.}(2017)Comisso, Lingam, Huang, \&
  Bhattacharjee}]{ComissoLHB2017}
---. 2017, submitted to Astrophys. J., arXiv:1707.01862

\bibitem[{Coppi {et~al.}(1976)Coppi, Galvao, Pellat, Rosenbluth, \&
  Rutherford}]{CoppiGPRR1976}
Coppi, B., Galvao, E., Pellat, R., Rosenbluth, M.~N., \& Rutherford, P.~H.
  1976, Sov. J. Plasma Phys., 2, 533

\bibitem[{Daughton {et~al.}(2009)Daughton, Roytershteyn, Albright, Karimabadi,
  Yin, \& Bowers}]{DaughtonRAKYB2009}
Daughton, W., Roytershteyn, V., Albright, B.~J., {et~al.} 2009, Phys. Rev.
  Lett., 103, 065004

\bibitem[{Daughton {et~al.}(2011)Daughton, Roytershteyn, Karimabadi, Yin,
  Albright, Bergen, \& Bower}]{DaughtonRKYABB2011}
Daughton, W., Roytershteyn, V., Karimabadi, H., {et~al.} 2011, Nature Physics,
  7, 539

\bibitem[{{De Pontieu} {et~al.}(2014){De Pontieu}, {Title}, {Lemen}, {Kushner},
  {Akin}, {Allard}, {Berger}, {Boerner}, {Cheung}, {Chou}, {Drake}, {Duncan},
  {Freeland}, {Heyman}, {Hoffman}, {Hurlburt}, {Lindgren}, {Mathur}, {Rehse},
  {Sabolish}, {Seguin}, {Schrijver}, {Tarbell}, {W{\"u}lser}, {Wolfson},
  {Yanari}, {Mudge}, {Nguyen-Phuc}, {Timmons}, {van Bezooijen}, {Weingrod},
  {Brookner}, {Butcher}, {Dougherty}, {Eder}, {Knagenhjelm}, {Larsen},
  {Mansir}, {Phan}, {Boyle}, {Cheimets}, {DeLuca}, {Golub}, {Gates}, {Hertz},
  {McKillop}, {Park}, {Perry}, {Podgorski}, {Reeves}, {Saar}, {Testa}, {Tian},
  {Weber}, {Dunn}, {Eccles}, {Jaeggli}, {Kankelborg}, {Mashburn}, {Pust},
  {Springer}, {Carvalho}, {Kleint}, {Marmie}, {Mazmanian}, {Pereira}, {Sawyer},
  {Strong}, {Worden}, {Carlsson}, {Hansteen}, {Leenaarts}, {Wiesmann},
  {Aloise}, {Chu}, {Bush}, {Scherrer}, {Brekke}, {Martinez-Sykora}, {Lites},
  {McIntosh}, {Uitenbroek}, {Okamoto}, {Gummin}, {Auker}, {Jerram}, {Pool}, \&
  {Waltham}}]{DePontieuTL2014}
{De Pontieu}, B., {Title}, A.~M., {Lemen}, J.~R., {et~al.} 2014, Solar Physics,
  289, 2733

\bibitem[{Fermo {et~al.}(2010)Fermo, Drake, \& Swisdak}]{FermoDS2010}
Fermo, R.~L., Drake, J.~F., \& Swisdak, M. 2010, Phys. Plasmas, 17, 010702

\bibitem[{Furth {et~al.}(1963)Furth, Killeen, \& Rosenbluth}]{FurthKR1963}
Furth, H.~P., Killeen, J., \& Rosenbluth, M.~N. 1963, Phys. Fluids, 6, 459

\bibitem[{Guo(2017)}]{Guo2017}
Guo, L.-J. 2017, private communication

\bibitem[{Guo {et~al.}(2013)Guo, Bhattacharjee, \& Huang}]{GuoBH2013}
Guo, L.-J., Bhattacharjee, A., \& Huang, Y.-M. 2013, Astrophys. J. Lett., 771,
  L14

\bibitem[{Guzdar {et~al.}(1993)Guzdar, Drake, McCarthy, Hassam, \&
  Liu}]{GuzdarDMHL1993}
Guzdar, P.~N., Drake, J.~F., McCarthy, D., Hassam, A.~B., \& Liu, C.~S. 1993,
  Phys. Fluids B, 5, 3712

\bibitem[{Huang \& Bhattacharjee(2010)}]{HuangB2010}
Huang, Y.-M., \& Bhattacharjee, A. 2010, Phys. Plasmas, 17, 062104

\bibitem[{Huang \& Bhattacharjee(2012)}]{HuangB2012}
---. 2012, Phys. Rev. Lett., 109, 265002

\bibitem[{Huang \& Bhattacharjee(2013)}]{HuangB2013}
---. 2013, Phys. Plasmas, 20, 055702

\bibitem[{Huang \& Bhattacharjee(2016)}]{HuangB2016}
---. 2016, Astrophys. J., 818, 20

\bibitem[{Huang {et~al.}(2011)Huang, Bhattacharjee, \& Sullivan}]{HuangBS2011}
Huang, Y.-M., Bhattacharjee, A., \& Sullivan, B.~P. 2011, Phys. Plasmas, 18,
  072109

\bibitem[{Innes {et~al.}(2015)Innes, Guo, Huang, \&
  Bhattacharjee}]{InnesGHB2015}
Innes, D.~E., Guo, L.-J., Huang, Y.-M., \& Bhattacharjee, A. 2015, Astrophys.
  J., 813, 86

\bibitem[{Ji \& Daughton(2011)}]{JiD2011}
Ji, H., \& Daughton, W. 2011, Phys. Plasmas, 18, 111207

\bibitem[{Karpen {et~al.}(2012)Karpen, Antiochos, \& DeVore}]{KarpenAD2012}
Karpen, J.~T., Antiochos, S.~K., \& DeVore, C.~R. 2012, Astrophys. J., 760, 81

\bibitem[{Knoll \& Chac\'on(2006)}]{KnollC2006}
Knoll, D.~A., \& Chac\'on, L. 2006, Phys. Plasmas, 13, 032307

\bibitem[{Kulsrud(2005)}]{Kulsrud2005}
Kulsrud, R.~M. 2005, Plasma Physics for Astrophysics (Princeton University
  Press)

\bibitem[{Lapenta(2008)}]{Lapenta2008}
Lapenta, G. 2008, Phys. Rev. Lett., 100, 235001

\bibitem[{Leake {et~al.}(2013)Leake, Lukin, \& Linton}]{LeakeLL2013}
Leake, J.~E., Lukin, V.~S., \& Linton, M.~G. 2013, Phys. Plasmas, 20, 061202

\bibitem[{Leake {et~al.}(2012)Leake, Lukin, Linton, \& Meier}]{LeakeLLM2012}
Leake, J.~E., Lukin, V.~S., Linton, M.~G., \& Meier, E.~T. 2012, Astrophys. J.,
  760, 109

\bibitem[{Loureiro {et~al.}(2012)Loureiro, Samtaney, Schekochihin, \&
  Uzdensky}]{LoureiroSSU2012}
Loureiro, N.~F., Samtaney, R., Schekochihin, A.~A., \& Uzdensky, D.~A. 2012,
  Phys. Plasmas, 19, 042303

\bibitem[{Loureiro {et~al.}(2007)Loureiro, Schekochihin, \&
  Cowley}]{LoureiroSC2007}
Loureiro, N.~F., Schekochihin, A.~A., \& Cowley, S.~C. 2007, Phys. Plasmas, 14,
  100703

\bibitem[{Loureiro \& Uzdensky(2016)}]{LoureiroU2016}
Loureiro, N.~F., \& Uzdensky, D.~A. 2016, Plasma Phys. Control. Fusion, 58,
  014021

\bibitem[{Lynch {et~al.}(2016)Lynch, Edmondson, Kazachenko, \&
  Guidoni}]{LynchEKG2016}
Lynch, B.~J., Edmondson, J.~K., Kazachenko, M.~D., \& Guidoni, S.~E. 2016,
  Astrophys. J., 826, 43

\bibitem[{McKechan {et~al.}(2010)McKechan, Robinson, \&
  Sathyaprakash}]{McKechanRS2010}
McKechan, D. J.~A., Robinson, C., \& Sathyaprakash, B.~S. 2010, Class. Quantum
  Grav., 27, 084020

\bibitem[{Murphy {et~al.}(2013)Murphy, Young, Shen, Lin, \&
  Ni}]{MurphyYSLN2013}
Murphy, N.~A., Young, A.~K., Shen, C., Lin, J., \& Ni, L. 2013, Phys. Plasmas,
  20, 061211

\bibitem[{Ng \& Ragunathan(2011)}]{NgR2011}
Ng, C.~S., \& Ragunathan, S. 2011, in ASP Conference Series, Vol. 444,
  Numerical Modeling of Space Plasma Flows: ASTRONUM-2010, ed. N.~V. Pogorelov,
  E.~Audit, \& G.~P. Zank (Astronomical Society of the Pacific), 124--129

\bibitem[{Ni {et~al.}(2012)Ni, Ziegler, Huang, Lin, \& Mei}]{NiZHLM2012}
Ni, L., Ziegler, U., Huang, Y.-M., Lin, J., \& Mei, Z. 2012, Phys. Plasmas, 19,
  072902

\bibitem[{Oishi {et~al.}(2015)Oishi, Low, Collins, \& Tamura}]{OishiLCT2015}
Oishi, J.~S., Low, M.-M.~M., Collins, D.~C., \& Tamura, M. 2015, Astrophys. J.
  Lett., 806, L12

\bibitem[{Parker(1957)}]{Parker1957}
Parker, E.~N. 1957, J. Geophys. Res., 62, 509

\bibitem[{Priest \& Forbes(2000)}]{PriestF2000}
Priest, E.~R., \& Forbes, T. 2000, Magnetic reconnection : {MHD} theory and
  applications (Cambridge University Press)

\bibitem[{Pucci \& Velli(2014)}]{PucciV2014}
Pucci, F., \& Velli, M. 2014, Astrophys. J. Lett., 780, L19

\bibitem[{Rutherford(1973)}]{Rutherford1973}
Rutherford, P.~H. 1973, Phys. Fluids, 16, 1903

\bibitem[{Schindler(2007)}]{Schindler2007}
Schindler, K. 2007, Physics of Space Plasma Activity (Cambridge University
  Press)

\bibitem[{Shen {et~al.}(2013)Shen, Lin, Murphy, \& Raymond}]{ShenLMR2013}
Shen, C., Lin, J., Murphy, N.~A., \& Raymond, J.~C. 2013, Phys. Plasmas, 20,
  072114

\bibitem[{Shepherd \& Cassak(2010)}]{ShepherdC2010}
Shepherd, L.~S., \& Cassak, P.~A. 2010, Phys. Rev. Lett., 105, 015004

\bibitem[{Shibata \& Tanuma(2001)}]{ShibataT2001}
Shibata, K., \& Tanuma, S. 2001, Earth Planets Space, 53, 473.
\newblock \url{www.terrapub.co.jp/journals/EPS/abstract/5306/53060473.html}

\bibitem[{Sironi {et~al.}(2016)Sironi, Giannios, \& Petropoulou}]{SironiGP2016}
Sironi, L., Giannios, D., \& Petropoulou, M. 2016, Monthly Notices of the Royal
  Astronomical Society, 462, 48

\bibitem[{Sweet(1958)}]{Sweet1958a}
Sweet, P.~A. 1958, in Electromagnetic phenomena in cosmical physics, ed.
  B.~Lehnert, Vol.~6 (Cambridge University Press), 123--134

\bibitem[{Tajima \& Shibata(1997)}]{TajimaS1997}
Tajima, T., \& Shibata, K. 1997, Plasma Astrophysics (Addison Wesley)

\bibitem[{Takamoto(2013)}]{Takamoto2013}
Takamoto, M. 2013, Astrophys. J., 775, 50

\bibitem[{Tenerani {et~al.}(2015)Tenerani, Velli, Rappazzo, \&
  Pucci}]{TeneraniVRP2015}
Tenerani, A., Velli, M., Rappazzo, A.~F., \& Pucci, F. 2015, Astrophys. J.
  Lett., 813, L32

\bibitem[{Uzdensky \& Loureiro(2016)}]{UzdenskyL2016}
Uzdensky, D.~A., \& Loureiro, N.~F. 2016, Phys. Rev. Lett., 116, 105003

\bibitem[{Uzdensky {et~al.}(2010)Uzdensky, Loureiro, \&
  Schekochihin}]{UzdenskyLS2010}
Uzdensky, D.~A., Loureiro, N.~F., \& Schekochihin, A.~A. 2010, Phys. Rev.
  Lett., 105, 235002

\bibitem[{Yamada {et~al.}(2010)Yamada, Kulsrud, \& Ji}]{YamadaKJ2010}
Yamada, M., Kulsrud, R., \& Ji, H. 2010, Rev. Mod. Phys., 82, 603

\bibitem[{Zweibel \& Yamada(2009)}]{ZweibelY2009}
Zweibel, E.~G., \& Yamada, M. 2009, Annu. Rev. Astron. Astrophys., 47, 291

\bibitem[{Zweibel \& Yamada(2016)}]{ZweibelY2016}
---. 2016, Proc. R Soc. A, 472, 20160479

\end{thebibliography}
\end{document}